%% file: paper.tex
\documentclass[twocolumn,10pt,twoside]{IEEEtran}

\usepackage[paperheight=11in,paperwidth=8.5in]{geometry}
\usepackage{amssymb}
\usepackage{arcs}
\usepackage{color}
\usepackage{geometry}
\usepackage{graphicx}
\usepackage{latexsym}
\usepackage{psfrag}
\usepackage{tikz}
\usepackage{xcolor}
\usepackage{algorithm}
\usepackage{algorithmic}
\usepackage{float}
\usepackage{wrapfig}
\usepackage{subfigure}
\newcommand{\kETAL}    {{\em et~al.}}

\newcommand{\bZ} {{\mathbb{Z}}}
\newcommand{\bG} {{\mathbb{G}}}

\floatname{algorithm}{Protocol}

\makeatletter
\def\ps@headings{%
\def\@oddhead{\mbox{}\scriptsize\rightmark \hfil \thepage}%
\def\@evenhead{\scriptsize\thepage \hfil \leftmark\mbox{}}%
\def\@oddfoot{}%
\def\@evenfoot{}}
\makeatother
\usepackage{amssymb}
\usepackage{amsmath}
\usepackage{algorithm}
\usepackage{algorithmic}
\usepackage{subfigure}
\usepackage{graphicx}
\usepackage{latexsym}
\usepackage{psfrag}

\def\BibTeX{{\rm B\kern-.05em{\sc i\kern-.025em b}\kern-.08em
    T\kern-.1667em\lower.7ex\hbox{E}\kern-.125emX}}
\makeatletter
\@addtoreset{equation}{section}
\makeatother
\renewcommand{\thesection}{\arabic{section}}
\renewcommand{\thesubsection}{\thesection.\arabic{subsection}}
\renewcommand{\theequation}{\thesection.\arabic{equation}}
\newtheorem{theorem}{\bf Theorem}      [section]

\newtheorem{definition}{\bf Definition}   [section]

\newcommand{\kENDdef}   {$\Box$}		
\newcommand{\cS}	{{\mathcal S}}
\newcommand{\cR}	{{\mathcal R}}

\begin{document}
\vspace{-0.2in}
\title{\huge AccConF: An Access Control Framework for Leveraging In-Network Cached Data in ICNs}
\author{Satyajayant Misra$^\dagger$, Reza Tourani$^\dagger$, Frank Natividad$^\dagger$, Travis Mick$^\dagger$, Nahid Ebrahimi Majd$^\ddagger$ and Hong Huang$^\star$\\
$^\dagger$ Computer Science Department, New Mexico State University, Las Cruces, New Mexico\\
Email:\{{\it misra, rtourani, fnativid,tmick}\}@cs.nmsu.edu\\
$^\ddagger$ Computer Science Department, California State University, San Marcos, California\\
Email:\{{\it nmajd}\}@csusm.edu\\
$^\star$ Electrical and Computer Engineering Department, New Mexico State University, Las Cruces, New Mexico\\
Email:\{{\it hhuang}\}@nmsu.edu\\
~\thanks{This work has been submitted to IEEE Transactions on Information Forensics and Security journal and is 
supported in part by the U.S. NSF grants:1345232 and 1248109 and the U.S. DoD/ARO grant: W911NF-07-2-0027.}
}
\maketitle
\pagestyle{empty}
\thispagestyle{empty}

\begin{abstract}
The fast-growing Internet traffic is increasingly becoming content-based and driven by mobile users, with users more 
interested in data rather than its source. 
This has precipitated the need for an {\em information-centric} Internet architecture. 
Research in information-centric networks (ICNs) have resulted in novel architectures, e.g., CCN/NDN, DONA, and PSIRP/PURSUIT;
all agree on named data based addressing and pervasive caching as integral design components.
%
%
%
With network-wide content caching, enforcement of content access control policies become non-trivial.
Each caching node in the network needs to enforce access control policies with the help of the content provider.
This becomes inefficient and prone to unbounded latencies especially during provider outages. 

In this paper, we propose an efficient access control framework for ICN, which allows legitimate users 
to access and use the cached content directly, and does not require verification/authentication 
by an online provider authentication server or the content serving router.
%
%
%
This framework would help reduce the impact of system down-time from server outages and reduce delivery latency 
by leveraging caching while guaranteeing access only to legitimate users. 
Experimental/simulation results demonstrate the suitability of this scheme for all users, but particularly for mobile users, 
especially in terms of the security and latency overheads. 

\end{abstract}


\input{sec01}
\input{sec02}

\input{sec03}
\input{sec04}
\input{sec05}


\vspace{-0.0in}
\section{Conclusions and Future Work}
\label{sec06}
In this paper, we present a novel access control framework (AccConF) for secure content delivery to legitimate users in ICNs.
Leveraging broadcast encryption, AccConF targets the users with power-constrained devices to enable efficient content access 
without involving an online authenticator. 
We detailed the protocols and the design decisions for the framework in the CCN/NDN architecture and demonstrated it's 
feasibility and scalability with practical experiments.
Our experimental/simulation results demonstrate that AccConF is practical and deployable with minimal network changes. 
It can be used by content providers to reduce latency and guarantee high availability of content. 

In future, we will optimize our smartphone application and the protocols, testing them in a large network. 
We will investigate more efficient system re-initialization when the system reaches its capacity. 
%
%

\vspace{-0.1in}
\bibliographystyle{plain}
\bibliography{paper}
%
%
\vspace{-0.5in}
\begin{IEEEbiography}[{\includegraphics[width=1in,height=1.25in,clip,keepaspectratio]{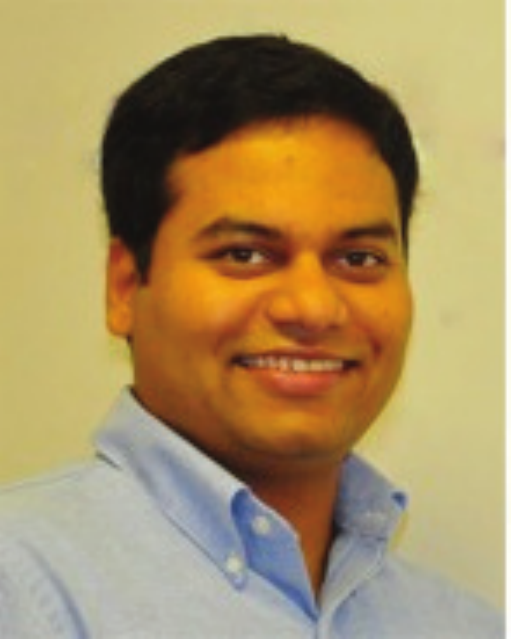}}]{Satyajayant Misra}
(SM'05, M'09) is an associate professor in computer science at New Mexico State University. He completed his M.Sc. in Physics and Information Systems from BITS, Pilani, 
India in 2003 and his Ph.D. in Computer Science from Arizona State University, Tempe, AZ, USA, in 2009. His research interests include wireless networks and the Internet, 
supercomputing, and smart grid architectures and protocols. He has served on several IEEE journal editorial boards and conference executive committees (Communications on Surveys 
and Tutorials, Wireless Communications Magazine, SECON 2010, INFOCOM 2012). He has authored more than 45 peer-reviewed IEEE/ACM journal articles and conference proceedings. 
More information can be obtained at www.cs.nmsu.edu/~misra. 
\end{IEEEbiography}
%
\vspace{-0.5in}
\begin{IEEEbiography}[{\includegraphics[width=1in,height=1.25in,clip,keepaspectratio]{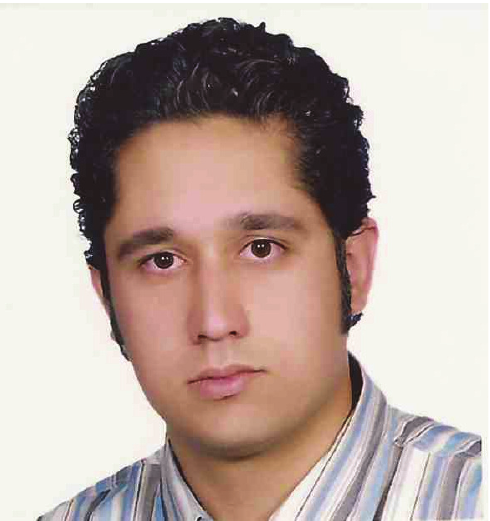}}]{Reza Tourani}
received his B.S. in computer engineering from IAUT, Tehran, Iran, in 2008, and M.S. in computer science from New Mexico State University, Las Cruces, NM, USA, in 2012. 
From 2013, he started his Ph. D. at New Mexico State University. His research interests include smart grid communication architecture and protocol, wireless protocols 
design and optimization, future Internet architecture, and privacy and security in wireless networks.
\end{IEEEbiography}
%
\vspace{-0.5in}
\begin{IEEEbiography}[{\includegraphics[width=1in,height=1.25in,clip,keepaspectratio]{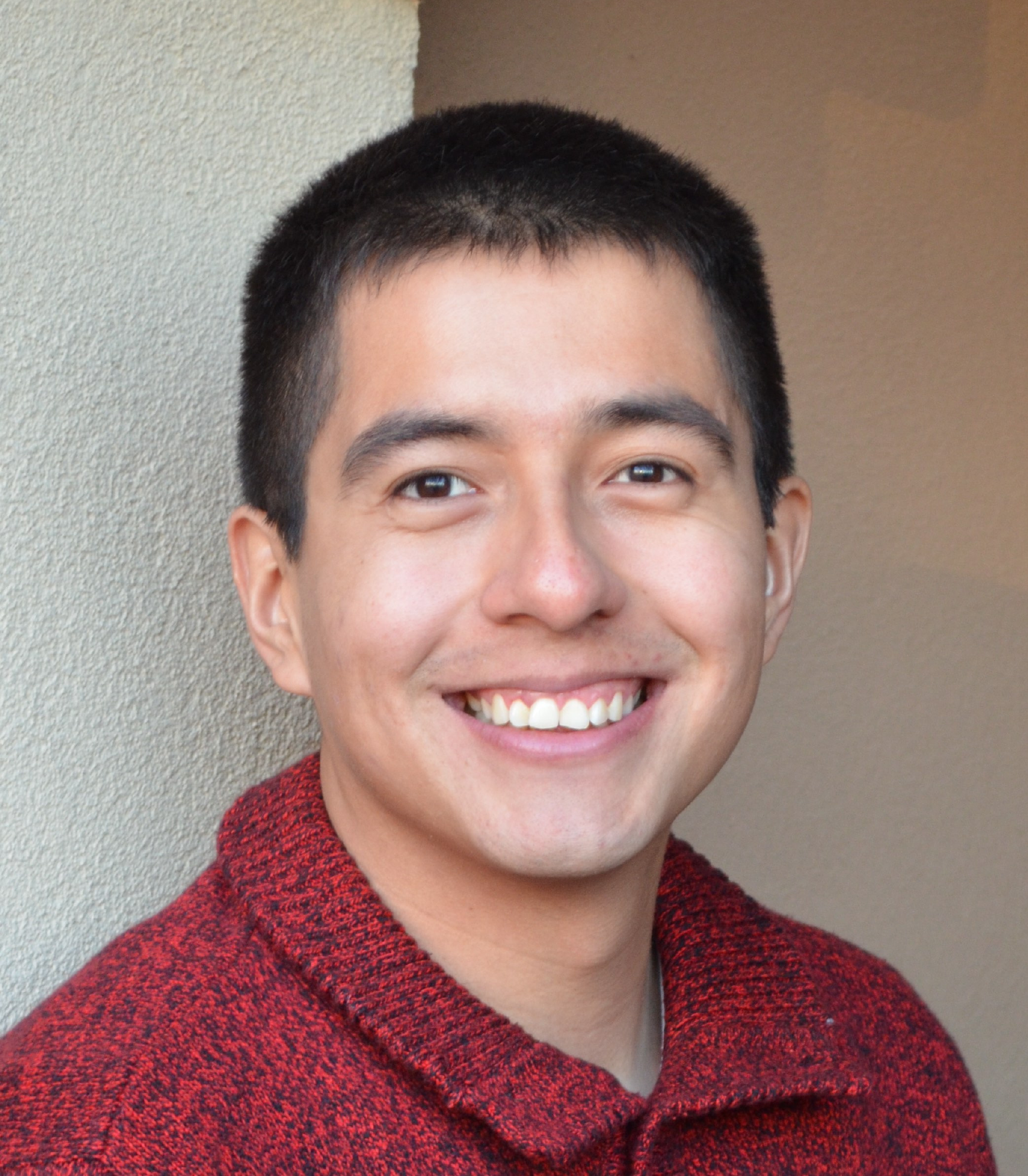}}]{Frank Natividad}
is currently pursuing his Master degree in the computer science department at the New Mexico State University, Las Cruces, NM, USA. Frank's current interests in research 
are in power trading agent competitions and machine learning in smart grid. 
\end{IEEEbiography}
%
\vspace{-0.5in}
\begin{IEEEbiography}[{\includegraphics[width=1in,height=1.25in,clip,keepaspectratio]{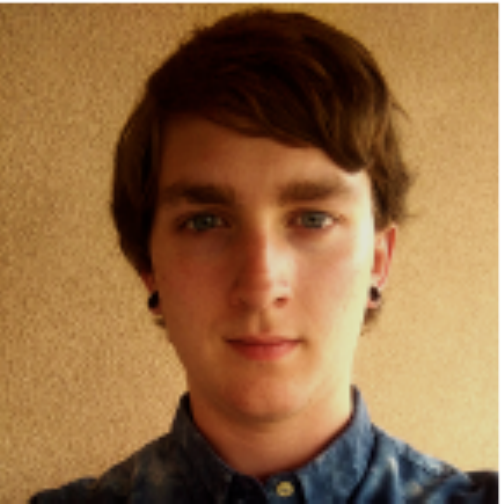}}]{Travis Mick}
completed his B.S. at New Mexico State University, Las Cruces, NM, USA in 2014, and is now pursuing an M.S. in computer science at New Mexico State University. His
research is in smart grid communication and information-centric networking.
\end{IEEEbiography}
%
\vspace{-0.5in}
\begin{biography}[{\includegraphics[width=1in,height=1.25in,clip,keepaspectratio]{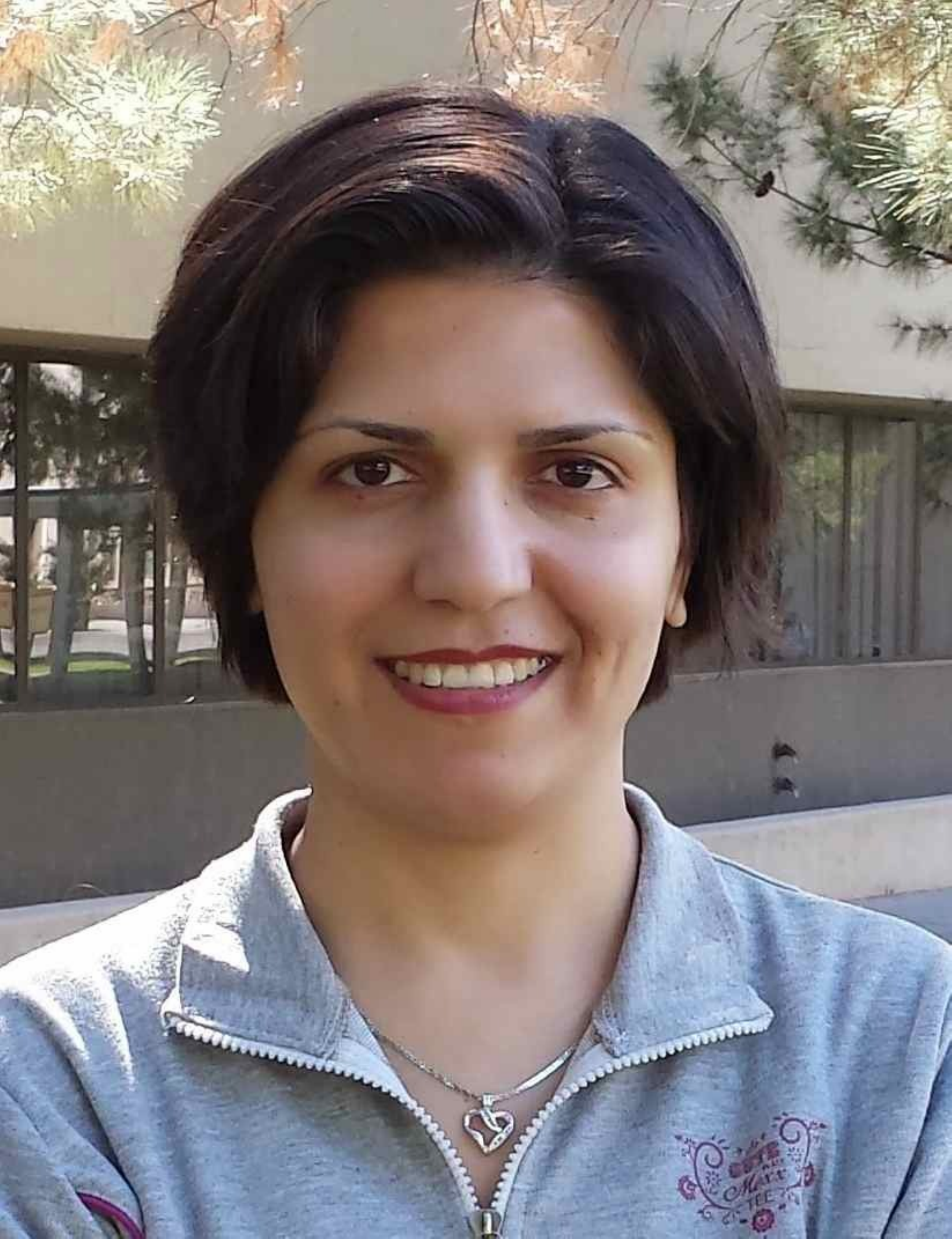}}]{Nahid Ebrahimi Majd}
received her PhD degree from the department of Computer Science, New Mexico State University, 
Las Cruces, NM, USA, in 2014. She is currently an assistant professor with the computer science department at the California State University at San Marcos. 
Her research interest is in energy harvesting wireless ad hoc networks, including relay node placement problem and cooperative caching problem in such networks.
\end{biography}
\vspace{-0.5in}
\begin{IEEEbiography}[{\includegraphics[width=1in,height=1.25in,clip,keepaspectratio]{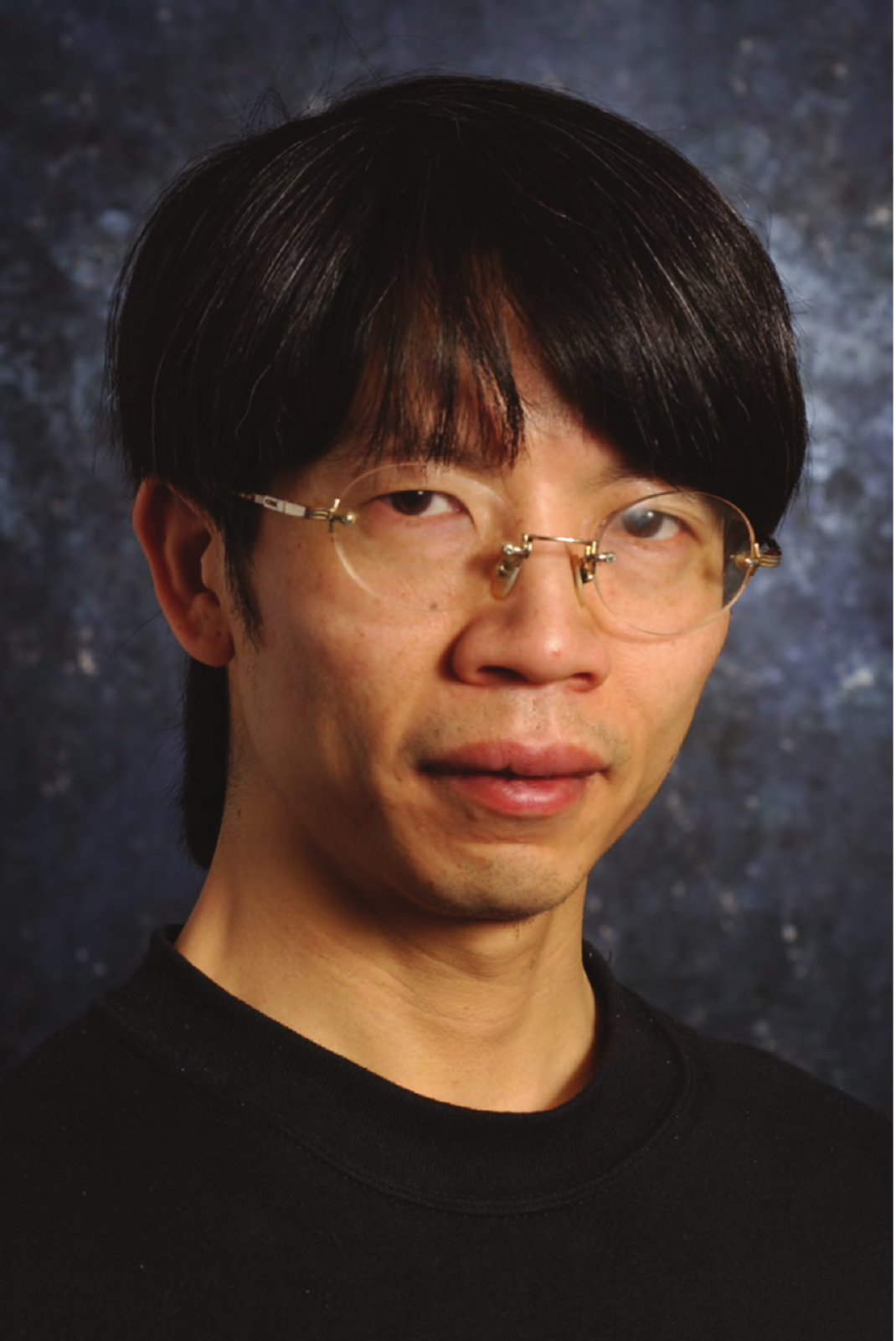}}]{Hong Huang}
received his B.E. degree from Tsinghua University, Beijing, China, and M.S. and Ph.D. degrees from Georgia Institute of Technology
in 2000 and 2002, respectively, all in electrical engineering. He is currently an associate professor
with the Klipsch School of Electrical and Computer Engineering at the New Mexico State University.
His current research interests include wireless sensor networks, mobile ad hoc networks, network security, and optical networks. He is a member of the IEEE.
\end{IEEEbiography}
\end{document}

%% file: sec01.tex
\section{Introduction}
\label{sec:01}%
The nature of the traffic and the service requirements from the Internet have changed tremendously. 
As per the Cisco Visual Networking Index Forecast (2019)~\cite{Cis2}: high bandwidth video traffic 
would account for $77\%$ of the Internet traffic by 2019 and mobile wireless devices will account for 
77\% of the world Internet traffic. 
This implies that the majority of the traffic on the Internet will be multimedia and emanate from 
wireless mobile users.  
This rapid growth has also been fueled by the use of P2P software (Ares, BitTorrent, etc.), 
which allows each user on the Internet to become a data server. 
This phenomenon has led to the Internet users becoming indifferent about 
the data source (video, music, movies) as long as they are reasonably sure about the content. 
These are alarming signs---the Internet was not engineered to scale for such trends. 
%

To address these concerns there has been a strong push to redesign the Internet architecture. 
This push is aimed at a shift from the {\it host-centric} Internet to the {\it information-centric} network~\cite{JacSmeTho09} where 
each data item is named and routing is performed using the name.  
The ICN Internet leverages pervasive in-network data caching and has built-in intelligence to satisfy requests by obtaining the data from 
network caches or the content provider, and transferring it to the requester(s). 
Several newly proposed Information-Centric Network (ICN) architectures, such as the 
CCN/NDN~\cite{JacSmeTho09}, DONA~\cite{KopChaChu07}, PSIRP~\cite{TarAinVis09}, PURSUIT~\cite{FotNikTro10}, 
and NetInf~\cite{Dan09}, aim to attain the above objectives.
We refer the interested readers to a survey on Information-Centric Networks~\cite{AhlDanImb12} for more information.

\begin{figure}
\centering
\includegraphics[width=3.2in]{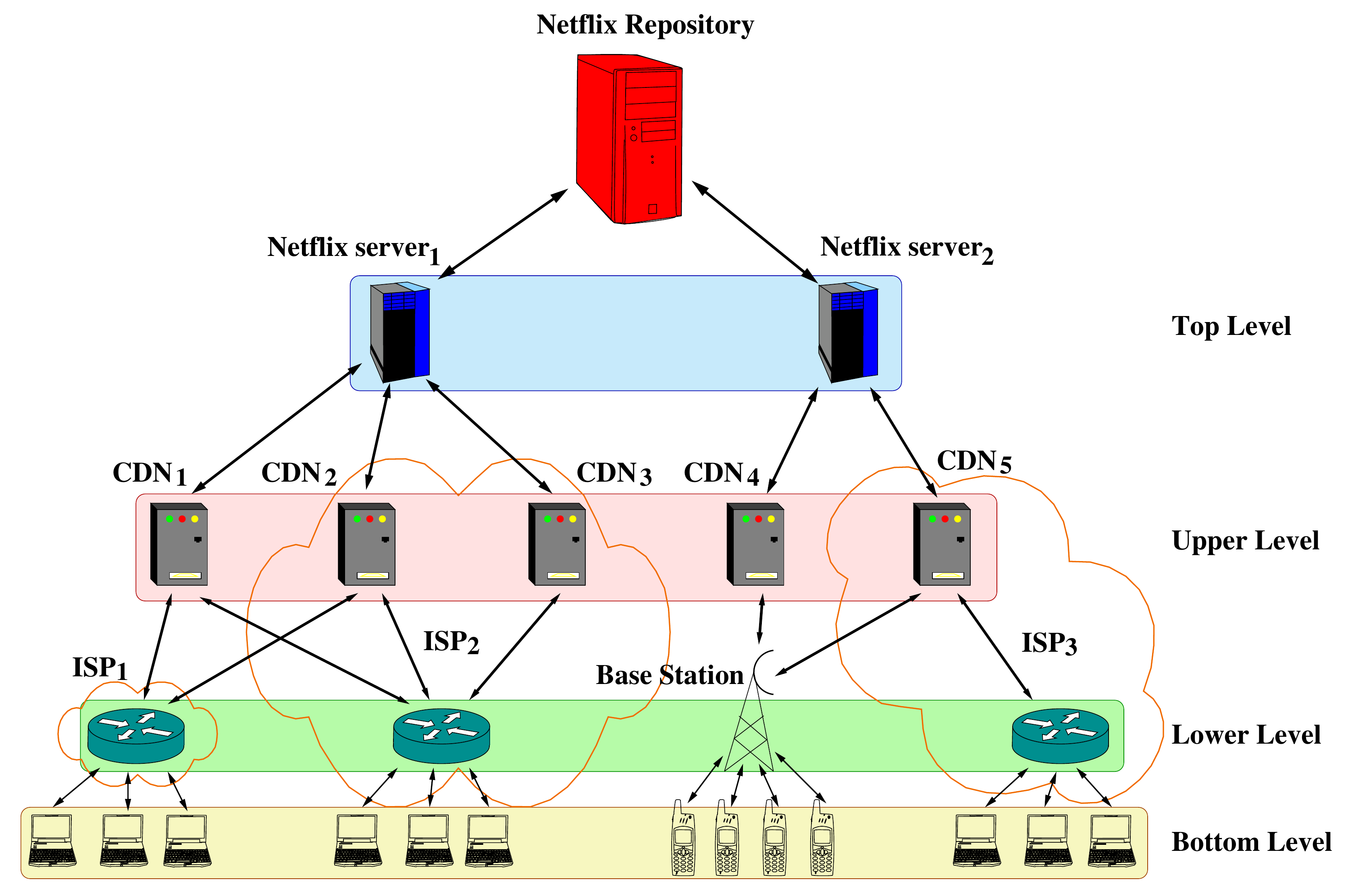}
\vspace{-0.2in}
\caption{Multi-level network architecture for Internet-based content distribution.}
\label{fig01}
\vspace{-0.24in}
\end{figure}

In today's Internet most Content Providers (CPs) use content distribution networks (CDNs) to cache (store) content 
geographically closer to the users for faster content delivery.
As shown in Fig.~\ref{fig01}, the Internet hierarchy consists of CPs at the top, followed by the CDNs (e.g. Akamai and Limelight), 
and then the ISPs (e.g. Comcast, AT\&T, and Verizon), culminating in the static/mobile end-users.
%
%
This architecture places most of the CDN nodes at the edge of ISPs (refer Fig.~\ref{fig01}) to reduce the network traffic; 
yet the ISPs keep deploying more network resources to handle the explosive data growth. 
The ICN paradigm, with its decoupling of data from the source, will enable in-network caching by the ISPs, 
reducing their network traffic load and improving scalability and data availability~\cite{Wan11}. 
{\em But, the important concern is how to ensure that the available cached content are only usable by authentic/legitimate users?}   

Let's illustrate this concern using Netflix as the CP and the CCN/NDN Internet architecture~\cite{JacSmeTho09}. 
To ensure user authenticity, in the current architecture, a legitimate user's Netflix player authenticates itself to a server 
hosted on a Cloud service (e.g., Amazon EC2).
Once the server authenticates the user, the player/client connects to a CDN node (selected based on network load, proximity, etc.) to access the content.
The access control (AC) is enforced by the server and subsequently, streaming happens from the designated CDN node. 

With ICN, ubiquitous caching  would require each node that caches any portion of a content to enforce the AC policies; an impractical exercise.
%
%
%
To cope with this problem, the user still has to authenticate himself to Netflix.
The decryption key, for the encrypted cached content, is granted to the user upon successful authentication.
%
%
%
However, there is an obvious concern; in our illustration, if the cloud service, Amazon EC2, is down, then the Netflix service is down.
The user cannot authenticate himself to use the cached content.
This has occurred several times in the past. 
%
One may argue that this service-loss can be addressed through better service-level agreements (SLAs) 
with the cloud provider, but even the best SLA cannot guarantee zero downtime. 
%
%
A better approach is one that can leverage the data available in routers close to the 
users, to satisfy requests from legitimate users. 
%
%

This research is {\bf motivated} by these observations. 
We address the question: {\it Can we design an efficient AC framework to utilize 
the cached content in ICNs that only serves legitimate users/subscribers?} 
In this paper, we extend our preliminary framework to answer this question~\cite{MisTouMaj13} (ACM ICN Workshop, 2013) and show that our framework 
also increases content availability (even when the provider's authentication service is offline) and improves clients' quality of experience.
In a nutshell, our {\bf contributions} include: {\bf {\em (i)}} Design of {\textit {\bfseries AccConF}} a {\em novel} ICN AC 
framework to guarantee trusted content in nearby caches can be efficiently used by only {\em legitimate users/subscribers}. 
AccConF leverages broadcast encryption and specifically targets mobile users that are at the low-end of the devices 
capability/power spectrum. 
Our framework also obviates the need for an ``always online'' authenticator/verifier.  
{\bf {\em (ii)}} Discussions on design and implementation issues of AccConF in the popular CCN/NDN architecture. 
{\bf {\em (iii)}} Proof that AccConF can handle user revocations limited by a large threshold $t$ and can be 
augmented to handle more than $t$ revoked users. 
{\bf {\em (iv)}} Implementation of AccConF in a CCN/NDN testbed and the ndn-SIM simulator on ns-3 and accompanying analysis 
validating its usability in mobile devices. 

In Section~\ref{sec01a}, we present the related work. 
In Section~\ref{sec02}, we present the basic definitions and notations, and in Section~\ref{sec03}, 
we present the system model, security assumptions, and the threat model. 
We present our framework in Section~\ref{sec04}, its ICN specific details in Section~\ref{sec05a}, 
and discuss its security provisions in Section~\ref{sec05b}. 
In Section~\ref{sec05}, we present our experimental results and analysis. 
In Section~\ref{sec06}, we present our conclusions. 

\vspace{-0.1in}
\section{Related Work} 
\label{sec01a} 
%
%
%
%
In CCN/NDN~\cite{JacSmeTho09}, the user's data interest (request) is either served by an intermediate router that receives the interest 
and has the data cached or the Content Provider (CP). 
Data is routed back using information stored in a router's pending interests table and the forwarding information base, 
and is cached at each forwarding router. 
In DONA~\cite{KopChaChu07}, CPs advertise their named content, in form of P:L where P is the hash of their public key and {\bf L} 
is the content's unique label, to resolution handlers (RHs), which form an inter-domain RH-hierarchy. 
A user transmits a data request with the help of the RH-hierarchy to a data source, which then transmits the data back along the 
same path. 
The data can be cached in the buffer of the involved RHs along the return path. 
The design paradigms of both PURSUIT~and~PSIRP~\cite{FotNikTro10} involve three separate elements -- 
publishers, subscribers, and the REndezvous NEtwork (RENE) with similar naming scheme as DONA. 
Rendezvous Points (RPs) in RENE perform rendezvous action between publishers and subscribers and select a path for a publisher/subscriber pair.
%
%
Network of Information (NetInf)~\cite{Dan09} provides a service conceptually similar to 
the rendezvous service in PSIRP/PURSUIT. 
%
{\em Caching and named data based addressing are integral facets of all these architectures, which are also the only two pre-requisites of 
our framework.}  

AC in the ICN has recently received more attention from the community~\cite{AriKop12, CheLeiXu14, FotMarPol12, GhaSchTsu15, IonZhaSch13, LiVerHua14, LiZhaZhe15}.
In~\cite{AriKop12}, the authors proposed a per-user privacy design in which content chunks are mixed with chunks of cover 
and the results are published into the network.
The user gets the necessary decoding information via a secure back channel from the CP, which requires the CP to be always online.
%
%
%
Fotiou~\kETAL~proposed an AC enforcement delegation technique~\cite{FotMarPol12}.
This scheme introduces the Relaying Party (RP) and the Access Control Provider (ACP) entities, which are responsible for 
storing the content and enforcing the AC policies, respectively.
%
%
The RP (a caching node) receives the user's request and sends a secret and the corresponding AC policy to both the ACP and the user.
The user authenticates himself to the ACP by forwarding the received secret, the policy, and his credentials.
The ACP authenticates the user and notifies the RP to transfer data.
This technique requires interaction between each router and an ACP, which is not scalable. 
%

Chen~\cite{CheLeiXu14} proposed a probabilistic encryption-based AC which leverages symmetric/asymmetric cryptographic operations.
The proposed mechanism was augmented with a Bloom filter representing the authorized clients' public keys; that is used by the intermediate routers 
to verify a user before forwarding the encrypted content.
In~\cite{GhaSchTsu15}, the authors proposed a mechanism built upon name obfuscation and authorized disclosure; the former prevents the unauthorized 
clients to obtain the content name.
The latter requires any entity, with a copy of the content, to perform client authentication and authorization.
In this scheme, the content name is encrypted (hashed) to prevent unauthorized access; the content is in plaintext.

Li~\kETAL~\cite{LiZhaZhe15} designed a light-weight signature and AC enforcement mechanism that uses per-content tokens.
Tokens are generated and assigned by the content provider to the network entities according to their capabilities.
Legitimate users authorize themselves to a network router by obtaining the content's private token(s) from the content provider and verifying to the router that 
they have the token(s).
%
%
%
This scheme suffers from the storage overhead of the token (three per content) at the routers and the overhead of token synchronization which 
undermine scalability. 
Also, the mechanism does not scale in the face of user-revocation as it requires complete re-keying at all routers. 

Attribute-based AC has also been investigated for ICN~\cite{IonZhaSch13, LiVerHua14}.
In~\cite{IonZhaSch13} the authors proposed a sketch of the key-policy and the ciphertext-policy based AC.
In the key-policy, the content is encrypted with a key that is derived from the content attributes and the access policy is embedded in the decryption key.
For the ciphertext policy, the access policy includes the authorized clients' attributes which is used to generate the decryption key.
Li~\kETAL~\cite{LiVerHua14} proposed a ciphertext-policy scheme in which the provider encrypts the content with a symmetric key. 
It then encrypts the symmetric key with the access policy, which results in the content name.
The user first acquires the content name from the name publishing system.
Only an authorized user can decrypt the content's name using his attributes to get the symmetric key.
The problem with the attribute-based systems is again lack of support for client revocation and computation complexity. 
%
%
%
%
%

Broadcast encryption (BE) was first proposed by Fiat and Naor~\cite{FiaNao94} to enable a source to send encrypted 
data to a set of legitimate users in the network who can decrypt the data. 
The protocol was $t$-resilient (resilient to collusion of up to $t$ malicious users) with $O=(t^2 \log^2 t \log n)$ message 
transmission overhead and  $O=(t \log t \log n)$ key storage at the user, where $t$ is the revocation threshold and $n$ is the 
total number of users.
In $2001$, Naor~\kETAL~\cite{NaoNaoLot01} decreased the key storage requirement to $\log n$. 
Broadcast encryption has found use in the real-world applications. 
For instance, subset difference based BE is used for AACS, HD DVD, Pay Television, and Blu-ray disc encryption. 
However, these techniques are not readily usable for secure content delivery on power and computation 
constrained mobile devices. 
%
%
%
%
%
%

In this paper, we extend our preliminary work~\cite{MisTouMaj13}, which uses the public-key based traitor tracing $t$-resilient algorithm proposed 
by Tzeng and Tzeng in~\cite{TzeTze01} as a building block to create a secure content delivery framework especially applicable for mobile devices.
%
%
%
%
{\em Our enhanced framework (AccConF) ensures that mobile devices need less than $4$ additional seconds at start-up on account 
of the BE procedures.} 
This guarantees that user experience is not affected adversely.  
We also, propose a detailed protocol for handling $|\cR| > t$, and address the real-world implementation challenges and present more 
analysis and experimental results.  
%
%

Majority of the proposed AC mechanisms in ICN either need an entity (or a network of entities) for client authentication and/or require the intermediate 
routers to perform client authentication.
These assumptions, on one hand, undermine the scalability of the system due to the additional workload of the routers. 
On the other hand, they undermine the security; in case a router decides to maliciously authenticates an unauthorized client.
Different from prior work, in AccConF, there is no entity for AC enforcement; the intermediate routers only forward an extra content 
(enabling block), which is much smaller than the original content.
In the event of client revocation, our framework only incurs a minor updating cost as opposed to the proposed mechanisms, 
which invariably require system re-keying.
%
%
%
%

%% file: sec02.tex
\section{Basic Definitions and Notations}
\label{sec02}%
From here on, we denote the content provider as $CP$, the content distribution 
network as $CDN$, a CDN node $i$ as $CN_i$, the Internet Service Provider as $ISP$. 
In our framework, the CP and its servers are essentially the same as they perform the same tasks, 
so we use the terms server and CP interchangeably.
A user $u_i$'s public/private key pair for asymmetric encryption/decryption is denoted as $<P_i, Pr_i>$ and 
the CP's corresponding key pair is denoted as $<P_{\cS}, Pr_{\cS}>$. 
Now we define some key concepts used in the paper (please refer to~\cite{MenOorVan97} for details).

\begin{definition}
\label{defn01} 
{\bf [Broadcast Encryption]} 
Broadcast encryption is defined as a mechanism where a CP can securely broadcast 
content to a set of legitimate users ${\cal U} = \{u_1, u_2, \ldots, u_n\}$, such that each $u_i \in {\cal U}$ 
can decrypt the content using his private key (or share). ~\cite{NaoPin01}
\hfill\kENDdef      
\end{definition}
%
\begin{definition}
\label{defn02}
{\bf [Shamir's $(t+1, n)$-threshold Secret Sharing Scheme]}
In this secret sharing, a secret is shared between $n$ users in a way that at least 
$(t + 1) \leq n$ users have to combine their shares to obtain the secret. 
No combination of users less than $t+1$ ($t+1$ is termed the {\em threshold}), can decipher the secret. 
This scheme is implemented with the help of a one-dimensional $t$-degree polynomial 
$p_{t}(x) = a_0 + a_1 x + a_2 x^2 + \ldots + a_t x^t$, which can be uniquely determined using 
any $t+1$ points on the polynomial. 
A user $u_i$'s share is given by $(x_i, f(x_i))$, where $x_i$ is a point on the X-axis and $f(x_i) = p_t(x_i)$.  
%
%
In Shamir's secret sharing scheme, generally the secret is the term $a_0$ in the polynomial.  
\hfill\kENDdef      
\end{definition} 
\begin{definition}
\label{defn03}
{\bf [Lagrangian Interpolation Polynomial]} 
A Lagrange's polynomial of degree $n$ taking on the values $f(x_0), \ldots, f(x_n)$ for the points 
$x_0, \ldots, x_n$ is given by, 
\[ 
L_{n}(x) = f(x_0) \frac{(x - x_1)(x-x_2) \ldots (x - x_n)}{(x_0 - x_1)(x_0 - x_2) \ldots (x_0 - x_n)} + 
\] 
\[f(x_1) \frac{(x - x_0)(x-x_2) \ldots (x - x_n)}{(x_1 - x_0)(x_1 - x_2) \ldots (x_1 - x_n)} + \ldots + \]
\[f(x_n)  \frac{(x - x_0)(x-x_1) \ldots (x - x_{n-1})}{(x_n - x_0)(x_n - x_1) \ldots (x_n - x_{n-1})}.\]
Note that the secret $a_0$ in Shamir's secret sharing scheme can be obtained as $a_0 = L_{n}(0)$. 
In this paper, we denote the $i^{th}$ fractional term (also called the Lagrangian coefficient)  in $L_{n}(0)$ as,  
$\lambda_i = \Pi_{0 \leq j (\neq i) \leq n} \frac{x_j}{x_j - x_i}$ resulting in 
$a_0 = L_n(0) = f(x_0) \lambda_0 + f(x_1) \lambda_1 + \ldots + f(x_n) \lambda_n$.    
\hfill\kENDdef      
\end{definition}
With Shamir's secret sharing, when $t+1$ users combine their shares, they can obtain a unique interpolating polynomial 
using well-known techniques, such as Lagrangian interpolation. 
The Lagrangian interpolation method uses the Lagrangian Interpolation Polynomial to interpolate $p_t(x)$. 
%

\begin{definition}
\label{defn04}
{\bf [Decisional Diffie-Hellman (DDH) Problem]} 
Let $G$ denote a multiplicative finite cyclic group of order $Q$ (a large prime number), and $g$ be a 
generator of $G$, then given two distributions $(g^x, g^y, g^{xy})$ and $(g^x, g^y, g^z)$, where 
$x, y, z \in {\cal Z}_Q$ (${\cal Z}\setminus Q{\cal Z}$), the set of non-negative integers truncated 
by $Q$, and are chosen at random, can the two distributions be distinguished?   
This DDH problem is widely assumed to be intractable~\cite{MenOorVan97}. 
\hfill\kENDdef      
\end{definition}
%
%
%
\begin{definition}
\label{defn05}
{\bf [Schnorr Group]} 
Given two large primes $Q$ and $P$, where $P = rQ + 1$, $r \in \bZ^{*}_Q$, where $\bZ^{*}_Q$ is the multiplicative 
group of integers $mod$ $Q$, choose $1 < h < P$, such that $ h^r \not\equiv 1$ $mod$ $P$, then $g = h^r$ generates a Schnorr group, 
which is a subgroup of $\bZ^{*}_P$, the multiplicative group of integers $mod$ $P$ of order $Q$.~\cite{Sch91} 
\hfill\kENDdef      
\end{definition}
%
We use the Schnorr group for our framework. 
In cryptography, such prime-order subgroups are desirable as the modulus is as small as possible relative to $Q$.

%% file: sec03.tex
\vspace{-0.1in} 
\section{System and Threat Models and Assumptions}
\label{sec03} 
In this section, we present the system model, our security assumptions, and possible security threats to our 
framework. 
\vspace{-0.3in} 
\subsection{System Model}
\label{sec03-01}
%
As the next generation Internet architecture is a notion that is constantly in flux, we model AccConF to 
be adaptable. 
%
The setup is hierarchical as shown in Fig.~\ref{fig01}, where the CPs (or their servers) form the top level of the network. 
The servers may be synchronized to have the same global image of the user base or be distributed, serving non-overlapping user groups, while 
still having access to the central content repository. 
The content is pushed onto the CDN nodes--the next (upper) level of the system hierarchy--to 
transmit the data to the users connected to the ISPs.     
The next (lower) level consists of ISPs, which cache the packets and forward the data to the users (bottom level). 
%

For illustrating our framework and experimentation, we use the CCN/NDN architecture~\cite{JacSmeTho09}, and its code-base~\cite{Ccnx}.  
However, with all ICN architectures sharing the same premise of caching and name based routing our framework will apply to all. 
%
%
%
%
In our framework, $Q$ and $P=2Q+1$ are large prime numbers, $n$ is the number of users in the system, $t$ is the number of 
users that can be revoked without affecting system performance; and given that all polynomial operations happen in $\bZ^{*}_Q$, $n$ 
has to satisfy the condition $n \leq Q - t - 1$.
A first-time user registers with the CP to get his credentials and can obtain data from proximal nodes or the CP.
 
\vspace{-0.1in} 
\subsection{Set-up and Security Assumptions} 
\label{sec03-02} 
%
We assume that the content is encrypted by the content provider using a secure symmetric key encryption algorithm, 
such as AES~\cite{MenOorVan97}. 
A content or a group of contents (set of movies) may be encrypted using the same secret key--a 
legitimate user can decrypt the set of contents after successfully extracting the key.   
Different secret keys can be used by the provider to encrypt different contents or groups of contents;
this allows the provider to define diverse AC policies.
Our framework's objective is to ensure that the content is encrypted and cannot be 
used by an entity that is not a legitimate user/client (not even CDN/ISP nodes).   
%
%

We also assume that a legitimate user's front-end player does not store the symmetric key after decrypting the content, and that a user cannot 
tamper the player, which performs the decryption. 
Most content providers (Netflix, DirectTV, Comcast) have a player (set-top box, a standalone or 
embedded player), which performs the task of decryption of the content and these players are not easily tamperable.  
Without this assumption, no known encryption scheme can be used for security. 
%
%
We assume that the user does not use VPN tunneling or other location-cloaking mechanisms, such as the Tor network~\cite{Tor}, to 
hide their location. 
In the rest of the paper, we use the term {\it user} and {\it client} to refer interchangeably to the {\it user's mobile device}.  
  
\vspace{-0.15in} 
\subsection{Threat Model} 
\label{sec03-03} 
%
In a set-up for content delivery, data security is of utmost importance. 
The use of symmetric key infrastructure, public key infrastructure, and our 
framework guarantees data security. 
However, there are several other attack scenarios. 
For instance, an attacker could flood the network with fake interests (new or replayed interests), thus orchestrating a denial of service (DoS) attack.
%
An adversary can pollute the routers' caches by sending out unpopular requests~\cite{XieWidWan12}.   
Traffic analysis attack can be performed on a specific user to identify his content access pattern. 
A compromised or colluding user's keying materials can be extracted and used by an adversary, not part of the 
system, to gain access to the content by impersonation. 
The extracted keying materials can be used by an adversary to mount a Sybil attack~\cite{Dou02}. 
Also, few revoked users (popularly termed as traitors) can collude to generate a key for a malicious user (pirate),  
not part of the system, to decode the content. 
Additionally, there are standard attacks by an adversary, such as chosen plaintext attack (CPA), 
chosen ciphertext attack (CCA), and adaptive chosen ciphertext attack (A-CCA)~\cite{MenOorVan97,TzeTze01}. 
By virtue of the information-centric paradigm, our framework, can address most of the threats mentioned. 
For instance, the use of the sequence numbers in the interest and data packets, and caching at the edge routers 
can help neutralize replay attacks. 
Aggregation of interest packets and controlling interest rates will mitigate DoS attacks. 
Note that neither the NDN architecture nor our framework require the users to identify themselves 
to the communicating hosts nor in the interest packets. 
This ensures identity privacy, unless of course, the routers in the user's neighborhood collude to identify him. 
Cache pollution attacks has already been addressed satisfactorily~\cite{XieWidWan12}. 
After proposing our framework, we will discuss its security against Sybil, 
collusion, and the CPA, CCA, and A-CCA attacks.  
%

%% file: sec04.tex
\section{AccConF: Framework for High Availability and Efficiency in Secure Content Delivery}
\label{sec04}
%
Now, we present our framework, which helps perform the following for AC in an ICN:
(i) Allows ISPs to cache the content packets at their edge-routers enabling requests for same data 
to be served from the cache.
%
(ii) Increases the availability of the content to users by not requiring an initial authentication 
by an online server. 
(iii) Ensures that only legitimate users can use the content, according to the content access policy, and no revoked user 
can use the contents.     
%
The protocols in our framework are either implemented at the top or the bottom levels of the system hierarchy (Fig.~\ref{fig01}). 
%
%
%
There are several BE schemes in the literature~\cite{FiaNao94,NaoNaoLot01,TzeTze01} and our framework is 
generic enough to use any BE scheme, which can account for user revocation. 
However, for ease of illustration in this paper, we use a BE scheme proposed by Tzeng and Tzeng~\cite{TzeTze01}, which is a variant 
of Shamir's secret sharing scheme (Definition~\ref{defn02}). 

In~\cite{TzeTze01}, the threshold $t+1$ of Shamir's scheme helps define a revocation threshold of $t$--the 
threshold for the number of user revocations permitted without affecting data secrecy. 
Congruently, we assume $n$ legitimate users in the system and the number of revoked users ($|\cR|$) to be 
at most $t$ $(<< n)$; we also propose an enhancement to handle $|\cR| > t$.    
The BE scheme proposed by Tzeng and Tzeng was proved by them to be as hard as the DDH problem~\cite{TzeTze01}.
%
%
We augment the proposed BE scheme to allow accurate and efficient encryption of the content, and to ensure that 
contents can only be used by legitimate users, but not by the revoked users.
%
%
Table~\ref{table1} presents the notations used to describe our framework. 

\setlength{\textfloatsep}{0.01cm}
\begin{table}
\centering
\caption{Notations Used}
\label{table1}
\begin{tabular}{|c|p{2in}|}
 \hline
 	Notation & Description \\
 \hline
	$P$, $Q$ & Big prime numbers such that $P=2Q+1$\\
	$\bZ_{Q}^{*}, \bZ_{P}^{*}$ & Multiplicative groups of integers of
 	 order $Q$ and $P$ respectively \\
	$\bG_Q, \bG_P$ & Cyclic groups of order $Q$ and $P$ respectively \\
	$g$ & Generator of a sub-group of $\bG_P$ of order $Q$ \\
	$ZQrand()$ & Random number generator in $\bZ^{*}_Q$ \\
  	$a_{0}$ & Constant of $p_{t}(x)$\\
  	$t$ & Degree of polynomial $p_{t}(x)$ and the
	   revocation threshold \\
  	$\cR$ & Set of revoked users, $|\cR| \leq t$ \\
	$n$ & Total number of legitimate users \\
 	$\tau$ & Secret (Symmetric) key for data encryption \\
	$T_i = (x_i, f(x_i))$ & Tuple of user $u_i$ \\ 
	$T_r = (x_r, f(x_r))$ & Tuple of revoked user $u_r$ \\
	$||$ & Concatenation operator \\	
 \hline
	$p_t(x)$ & One-dimensional $t$-degree polynomial \\
	$f(x_i)$ & Evaluation of coordinate $x_i$ on $p_t(x) \in \bZ^{*}_Q$ \\
	$E$ & Server's share (Protocol~\ref{proto1})\\
	$E^e$ & Transformed server's share (Protocol~\ref{proto2}) \\
	$\mathcal{S}_C$ & Enabling block \\
	$\gamma$ & Encrypted symmetric key $\tau$ \\
	$\Lambda$ & Set of partial Lagrangian coefficients \\
	$\Upsilon$ & Secret (Symmetric) key composed of smaller keys concatenation \\	
	$\lambda_k$ & $k^{th}$ Lagrangian coefficient \\
	$\hat{\lambda}{_k}$ & $k^{th}$ partial Lagrangian coefficient \\
 \hline
\end{tabular}
\end{table}
%
\subsection{Overview of AccConF}
\label{sec04-00}
Our framework consists of three major steps: The first two steps are performed at the server and are related to encrypting $\tau$, the 
symmetric key for data encryption; only the last step is formed at the client.
In the {\bf first step}, the server generates a polynomial of degree $t$ and evaluates $n+t$~$(>> t)$ number of points on it. 
The server distributes $n$ of the evaluated points among the $n$ clients, one to each legitimate client, while it keeps $t$ of 
the remaining as its own shares.
%
In the {\bf second step}, the server generates the enabling block -- an essential metadata block, which contains the encrypted $\tau$,  
and is used by a client in the last step to extract $\tau$. 
The enabling block is forwarded to the routers similarly as content chunks and forms an integral part of the content.
In the {\bf third step}, a legitimate client extracts the encrypted $\tau$ from the enabling block by using his share. 
%
%
\subsection{Basic Protocols} 
\label{sec04_01}
We use a server ${\cal S}$ to illustrate the computations at the server(s) or the CP. 
The server ${\cal S}$ generates the polynomial $p_{t}(x)$ and calculates the tuple $T_i = (x_i, f(x_i))$ for each 
legitimate user $u_i$. 
Where it does not create confusion, in the context of the users, we use share and tuple interchangeably. 
In what follows, we use index $i$ to represent the users' shares and index $j$ to represent the server shares.

\subsubsection{{\bfseries Polynomial and Shares Generation}}
\label{sec04_01_01} 

Protocol~\ref{proto1} presents the procedure for generation of the polynomial $p_t(x)$ of degree $t$. 
In Line~1, the server generates the $t+1$ coefficients of $p_t(x)$.
\begin{algorithm}[h]
\caption{Generation of Polynomial/User Shares at the Server}
\label{proto1}
\begin{algorithmic}[1]
\REQUIRE {Values of $n < Q$ and $t$, a prime number $Q$, $ZQrand()$.}
\ENSURE {Generates a polynomial $p_{t}(x)$ with random coefficients $a_0, \ldots, a_{t}$ and the tuple $T_j$ for each user $u_j$.}   

\STATE Calculates $a_i = ZQrand()$, $i = 0$ to $t$.
\STATE Generates $p_{t}(x)$ using the $a_i$s. 
\STATE Calculates $x_j = ZQrand()$, $j = 0$ to $t-1$ and $x_j~\ne~x_k, 0 \leq j, k \leq t-1$.
	\COMMENT{Ensures $x_j$s are positive, unique, and not reused for clients} 
\STATE Calculates $f(x_j) = p_{t}(x_j) \in \bZ^{*}_Q$, $j = 0$ to $t-1$.

\STATE Obtains $E = E \cup (x_j, f(x_j))$, $j = 0$ to $t-1$.
	\COMMENT{Calculation of each legitimate client's share follows.}  
\STATE Calculates $x_i = ZQrand()$, $i = t$ to $n + t$, and $x_i~\ne~x_k, 0 \leq i, k < n + t$. 
\STATE Calculates $f(x_i) = p_{t}(x_i)\in \bZ^{*}_Q$, $i = t$ to $n+t$. 
\STATE Stores values $T_{i} = (x_i, f(x_i))$. \COMMENT{Tuple of user $u_i$}
\end{algorithmic}
\end{algorithm}
%
%
%
It then generates its shares by identifying $t$ random points (Lines~3-5) on $p_{t}(x)$ and the $n$ clients' shares using $n$ other points (Lines~6-7).
%
%
%
The dissemination of the users' share happens through the User Registration Protocol (Protocol~\ref{proto3}).
%
%
The CP encrypts the content using a shared symmetric key $\tau \in \bZ^{*}_Q$. 
%
%
A bigger key (say $128$-bit AES key) can also be handled;  we will discuss this in Protocol~\ref{proto2}.  
\begin{algorithm}[h]
\caption{Generation and Encryption of Enabling Block}
\label{proto2}
\vspace{0.15in}
\begin{algorithmic}[1]
\REQUIRE {Server's share $E$, $ZQrand()$, $g \in \mathbb{G}_Q$, $a_0$, data secret key $\tau \in \bZ^*_Q$.}
\ENSURE {Enabling Block $\mathcal{S}_C$}   
\STATE Calculates $r = ZQrand()$.\\
\STATE Obtains $\gamma = \tau g^{ra_{0}}$. 
\COMMENT{$ra_{0} \in \bZ^{*}_{Q}$ and $\tau g^{ra_{0}} \in \bZ^{*}_{P}$.}\\
\STATE Calculates $g^{r} \in \bZ^{*}_{P}$.\\
\STATE Calculates partial Lagrangian coefficients  $\Lambda = \{ \hat{\lambda}{_k} \mid \hat{\lambda}{_k} = \prod_{0 \le j \ne k < t} \frac{x_{j}}{x_{j}-x_{k}} \in \bZ^{*}_Q$ \}.\\
\STATE Calculates $E^e = \{\langle x_j, g^{rf(x_j)}\rangle | (x_j, f(x_j)) \in E\}$ for the $t$s server shares and $rf(x_j) \in \bZ^{*}_{Q}$ and $g^{rf(x_j)} \in \bZ^{*}_{P}$. \\
\STATE $\mathcal{S}_C = \langle \gamma, g^r, \Lambda , E^e \rangle$
\STATE Generates a timeout value ($TO$) for $\cS_C$. 
\STATE Sign $\mathcal{S}_C$ using the private key ($Pr_{\cS}$) of the server. 
\end{algorithmic} 
\end{algorithm}
\hspace{-0.5in}
\subsubsection{{\bfseries Generation and Encryption of Enabling Block}} 
\label{sec04_01_02} 
Protocol~\ref{proto2} deals with the generation of the enabling block, which enables the legitimate user 
to extract the secret key $\tau$, and is delivered to the user as one of the first content packets.
%
%
By generating a random number (Line~1), the server obtains the encrypted secret key ($\gamma$) using the field generator ($g$), 
polynomial constant ($a_0$), and the secret encryption key ($\tau$) in Line~2.
Line~3 shows the transformation of the group generator, $g$, by an exponentiation operation with the generated random number $r$.
In Line~4, the server calculates $\Lambda$ (partial Lagrangian coefficients), this precomputed $\Lambda$ is used at the client for 
calculating the complete Lagrangian coefficients needed for decryption. 
As we will show in Section~\ref{sec05} by comparing our framework (Global) with the standard approach in literature (GlobalNP), 
this partial precomputation step helps reduce the decryption time at the client {\it tremendously}. 
Thus, our framework is computation-heavy at the server side, which result in lightweight computations at the clients. 
%
%
%
%
%

%
In Line~5, the server calculates the transformed enabling block, obtained by raising $g$ to the power of $rf(x_j)$ $\forall f(x_j) \in E$. 
%
%
In Line~6, the server puts together the enabling block $\cS_C$. 
%
%
We will discuss the need for timeout (Line~7) and how to decide a value for $TO$ in the next subsection. 
The enabling block $\cS_C$ is signed by the server (Line~8) to guarantee provenance.
A bigger key (say $128$-bit key for AES) can be used by splitting the bigger key $\Upsilon$ into smaller sub-keys 
$\Upsilon = \{\tau_1 || \ldots || \tau_b || \ldots || \tau_m\}$, where each $\tau_b \in \bZ^{*}_Q$ and instead of sending $\gamma$, 
the server can send $\{\gamma_1 = \tau_1 g^{ra_0}, \ldots, \gamma_m = \tau_m g^{ra_0}\}$. 
%
The user will combine the split keys to regenerate $\Upsilon$.
This protocol's ${\cal O}(t)$ modular exponentiations dominate its running time. 
\begin{algorithm}[h]
\caption{User $u_i$'s Registration}
\label{proto3}
\begin{algorithmic}[1] 
\REQUIRE{User's registration credentials.}
\STATE User $u_i$ creates a verifiable profile and successfully enters the system.
\STATE Server securely transmits the user its public key, $P_{\cS}$, its digital certificate, the user's share $(x_i, f(x_i))$, and 
the expiration time (TO) of the share. 
\end{algorithmic} 
\end{algorithm}
\subsubsection{{\bfseries New User Registration}} 
\label{sec04_01_03} 
 
Protocol~\ref{proto3} deals with registration of a new user in our framework. 
%
%
%
%
%
For registration, a user $u_i$ sends a {\em registration} interest to the CP. 
The format for the user's name is: $/Netflix/Registration/Unique$ \\ $\_User\_ID$.  
This interest contains $u_i$'s other credentials, encrypted with $P_{\cS}$ and signed by $Pr_i$ ($u_i$'s private key).
The CP then replies to $u_i$ with a data packet containing $u_i$'s unique valid share encrypted with $P_i$. 
The reply is unicast from the CP to $u_i$ and is not cached at intermediate routers.  
Even if the data is cached by a malicious router communication secrecy cannot be undermined. 
%

\subsubsection{{\bfseries Secret Extraction at the User}} 
\label{sec04_01_04} 
Protocol~\ref{proto4} presents the procedure used by $u_i$ to extract the secret key ($\tau$) needs to decrypt the content. 
%
User $u_i$ verifies the signature of $\cS_C$ (Line~1) that he has obtained along with the content. 
As per Definition~\ref{defn03}, the $k^{th}$ Lagrangian coefficient $\lambda_k$ is defined as $\prod_{0 \le j \neq k \le t} \frac{x_{j}}{x_{j}-x_{k}}$, 
where $0 \leq j, k \leq t - 1$ represent the indices of the server shares and are in $\cS_C$; the $t^{th}$ fraction is $x_i$ obtained from user $u_i$. 
The server precomputed $\Lambda$ (part of $\cS_C$) is used to obtain complete Lagrangian coefficients, thus reducing the 
computation time at the client significantly.
User $u_i$ simply calculates the last term $(\frac{x_i}{x_{i}-x_{k}})$, to obtain the $k^{th}$ Lagrangian 
coefficient by performing only one multiplication $(\lambda_k = \hat{\lambda}{_k}$ $\cdot ( \frac{x_i}{x_{i}-x_{k}} ))$, 
instead of $t$ multiplications (in Line~2). 
{\em This precomputation enables the framework's use in mobile devices.} 

The following steps obtain the parameters used for the decryption.  
%
%
Line~3 calculates $\delta_1$--the multiplication of the shares ($g^{rf(x_k)}, \forall f(x_k) \in E^e$) in $\cS_C$ raised to their corresponding 
Lagrangian coefficients ($\lambda_k$). 
The client calculates the Lagrangian coefficient of his share in Line~4 and derives $\delta_2$ through the same procedure as Line~3 (in Line~5).
%
%
With $\delta_1$, $\delta_2$ and the encrypted symmetric key $\gamma$, in Line~6 the client calculates the secret $\tau$. 
If the secret is $\Upsilon$ (a bigger key), then it can be obtained by a minor extension to Protocol~\ref{proto4}: in Line~6, instead 
of calculating just $\tau$, the user calculates $\{\tau_1, \ldots, \tau_b, \ldots, \tau_m\}$, using the same operations, but 
using $\{\gamma_1, \ldots, \gamma_b, \ldots, \gamma_m\}$, to recreate $\Upsilon = \{\tau_1|| \ldots ||\tau_m\}$.   
Once the user $u_i$ extracts $\tau$, then she can decrypt the content. 
%
%
This protocol requires ${\cal O}(t)$ modular exponentiations--again, the bulk of the running time. 
We detail the effects of different values of $t$ in the next section.    
%
%
%
%
%
\begin{theorem} 
\label{thm04_01}
A legitimate user $u_i$ can use the enabling block $\cS_C$ and his own tuple $(x_i, f(x_i))$ and correctly decrypt 
the secret key $\tau$ using Protocol~\ref{proto4}.  
\hfill$\square$
\end{theorem}
\begin{proof}
%
Note that the constant term of the polynomial $p_t(x)$ can be calculated using $L_n(0)$, 
$a_0 = L_n(0) = f(x_0) \lambda_0 + f(x_1) \lambda_1 + \ldots + f(x_t) \lambda_t$, where the Lagrangian coefficient 
$\lambda_{k} = \prod_{0 \leq j \neq k \leq t} \frac {x_{j}} {x_{j}-x_{k}}$ as per Definition~\ref{defn03}. 
Consequently, $ra_{0} = r \sum_{k=0}^{t} (f(x_{k}) \cdot \lambda_{k})$, where $r$ is a large random number. 
%
%
By virtue of the fact that $g$ is the generator of the Schnorr group (subgroup of $\bZ^{*}_P$) of order $Q$, the next few steps follow.
%
\begin{eqnarray}
	\delta_1 * \delta_2 & = & \prod_{k=0}^{t-1} (g^{rf(x_{k})})^{\lambda_{k}} \cdot (g^{r})^{f(x_i) \lambda_i} 
	  (\makebox{\footnotesize {Lines~3\&5 Protocol~\ref{proto4}.}}) \nonumber \\
	& = & g^{r\sum_{k=0}^{t-1} (f(x_{k}) \cdot \lambda_{k})} \cdot g^{r f(x_i) \lambda_i}  \nonumber \\
	& = & g^{r \left[ (f(x_{0}) \lambda_{0}) + (f(x_{1}) \lambda_{1}) + \cdots + f(x_{t-1}) \lambda_{t-1} + (f(x_{i}) 
	\lambda_{i})\right] } \nonumber \\
	& = &  g^{ra_0} \nonumber
\end{eqnarray}	
Hence, $\tau \cdot g^{ra_0}/ \{\prod_{k=0}^{t - 1} (g^{rf(x_{k})})^{\lambda_{k}} \cdot (g^{r})^{f(x_i) \lambda_i}\} = \tau$ 
and the user can obtain the secret key.  
\end{proof}
\begin{algorithm}[!th]
\caption{Secret Extraction by User $u_i$}
\label{proto4}
\begin{algorithmic}[1]
\REQUIRE{${\mathcal S}_C$ and $T_i$, the share of $u_i$}
\ENSURE{Secret key $\tau$ for data decryption}
\STATE Verifies the signature of $\cS_C$ using $P_{\cS}$. 
\STATE Calculates Lagrangian coefficient \\ $\lambda_k = \hat{\lambda}{_k}$ $\cdot ( \frac{x_i}{x_{i}-x_{k}} ) \in \bZ^{*}_Q$, {\hspace*{2em}} ($\forall \hat{\lambda}{_k} \in \Lambda$)\\
\STATE Calculates $\delta_{1} = \prod_{0\leq k \leq t-1} (g^{rf(x_k)})^ {\lambda_{k}}$, where $\delta_{1} \in \bZ^{*}_{P}$ and $g^{rf(x_k)} \in E^e$ contained in $\cS_C$. \\
\STATE Calculates its Lagrangian coefficient $\lambda_i = \prod_{0 \le j \textless t} \frac{x_{j}}{x_{j}-x_{i}} \in \bZ^{*}_Q$, where $x_{i}$ is obtained from $T_i$.\\
\STATE Calculates $\delta_{2} = (g^{r})^ {f(x_{i})\lambda_{i}}$, where $f(x_{i})\lambda_{i} \in Z^{*}_{Q}$, $\delta_{2} \in Z^{*}_{P}$, and $g^{r} \in \mathcal{S}_C$. \\
\STATE Extracts secret key $ \tau = \frac{\gamma}{\delta_{1} * \delta_{2}}$.\\
\end{algorithmic} 
\end{algorithm}
%
%
%
%
\vspace{-0.05in}
\section{ICN-Specific Details of AccConF}
\label{sec05a}
%
We now discuss the compatibility of the framework with popular ICN architectures. 
%
%
In publish-subscribe based schemes, such as PURSUIT~\cite{FotNikTro10,TarAinVis09} and NetInf~\cite{Dan09}, 
the content's meta-information (number of packets, encoding, etc.) are published by the CP, whereas, in CCN/NDN or DONA, 
this information can be elicited by an interest packet sent to the network or the resolution handlers respectively. 
%
%
%
%
Our framework requires no extra messages or complexity in the network to leverage data-naming and caching. 
It can be implemented at every node in the network, including the rendezvous nodes in PURSUIT and the resolution handlers in DONA. 
Below we discuss the specific design details of our framework from the perspective of the popular CCN/NDN architecture. 
%
%

\begin{figure}[h] 
\centering
\includegraphics[height=1.7in, width=3.2in]{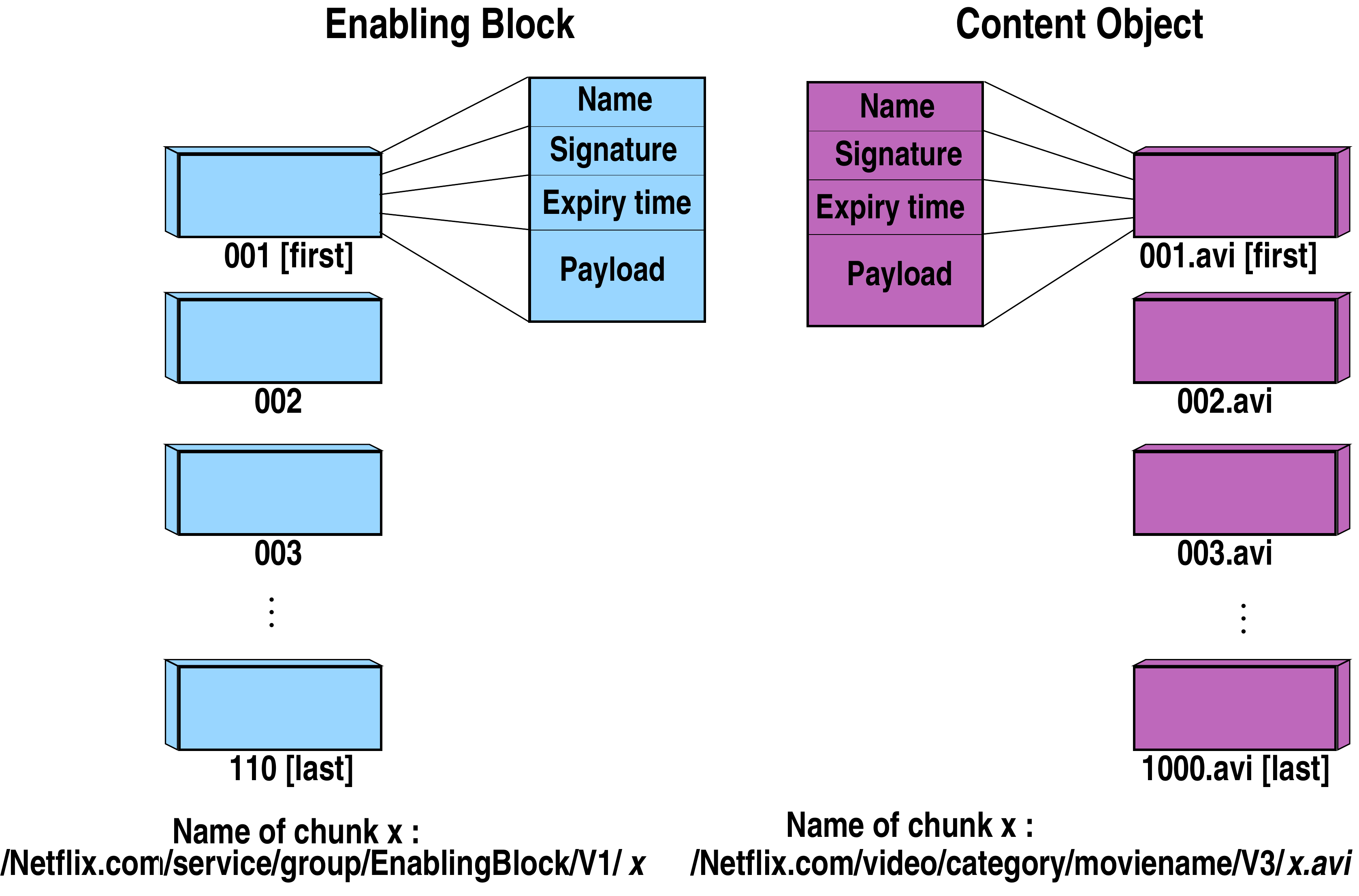}
\caption{Our naming scheme for the enabling block and content chunks.}
\label{fig02_02}
\end{figure}

\vspace{-0.08in}
\subsection{Data Chunking and Packet Naming}
\label{sec04_01a}
%
\subsubsection{Chunk Creation} 
%
Large contents are broken down into smaller data packets (chunks); each chunk is named uniquely and requested by 
its corresponding interest. 
%
%
%
%
%
Fig.~\ref{fig02_02} illustrates the splitting of the content and the enabling block. 
Both are split into equal sized chunks and given appropriate names for distinction. 
%

\subsubsection{Packet Naming} 
%
%
%
We follow the hierarchical naming convention of CCN/NDN (ref.~Fig.~\ref{fig02_02}). 
%
A typical content chunk name is {\em /Netflix.com/movie/category/movieName/V3/x.avi}.
The first segment is the CP's name, next (``movie'') is the content type,
followed by the category, e.g, Sci-Fi or Comedy, the fourth is the content name (Star Wars), 
the fifth is the version (V3), and the last part ($x.avi$) represents the chunk number.  
Versioning enables coexistence of different content qualities and expiry of content.

The enabling block naming follows the same convention but with the replacement of data type with the 
service type (premium, standard, plus). 
The category segment is replaced by the group with same intention--help group different users under the same service. 
Two types of numbering scheme can be used: sequential and random. 

%

\noindent \underline{Sequential Numbering:}
%
%
In this scheme, each content chunk has a sequence number $x \in \{001, 002, \ldots \}$, with increasing value of $x$.
This scheme is easy to implement, but enables cache probing and traffic analysis attack at the router/proxy~\cite{lauinger12}.

\noindent \underline{Random Numbering:} 
%
In random numbering, the value of $x$ for the first packet is known to the client; however, each subsequent packet has a random $x$ value. 
Each chunk carries the sequence number of the next interest to be used.
This helps negate traffic analysis attack but, may undermine aggregation of chunks. 
%

%


\vspace{-0.08in}
\subsection{Protocols to Handle System Dynamics}
\label{sec04_02}
%
Our framework has to address several system dynamics. 
%
For instance, {\it (i)} what to do when a registered user discontinues the service and needs to be revoked? 
{\it (ii)} What happens when the number of revoked users reaches the threshold $t$? 
{\it (iii)} What happens when a new user arrives and the system is at its capacity? 
We detail how these events are handled. 

\subsubsection{Revocation of a User $u_r$}
\label{sec04_02_01}   
%
When a user $u_r$ has to be revoked, the server replaces 
one of its $t$ tuples in $\cS_C$ with $T_r = (x_r, f(x_r))$,  $u_r$'s tuple. 
Hence in Line~5 of Protocol~\ref{proto2}, one of the $\langle x_j, g^{rf(x_j)} \rangle$ has to be replaced with $ \langle x_r, g^{rf(x_r)} \rangle$, 
thus changing $\cS_C$ to $\cS_C^\prime$. 
Several concerns that need to be addressed on this front are:  
(a) A high rate of revocation would require a new $\cS_C^\prime$ to be disseminated in the network with every revocation.  
Hence, the enabling block should be a small overhead and should be named in a way that allows differentiation between multiple versions 
in the network.
(b) The new $\cS_C^\prime$ has to be refreshed everywhere data exists, so that the revoked user cannot access the content. 
 
In Section~\ref{sec05}, using implementation results we show that the size of $\cS_C$ is much smaller than the 
content size ($<1\%$). 
Also if one key is used to encrypt several contents (e.g., movies), the amortized cost over all related contents can be made negligible. 
%
%
%
Thus (a) can be addressed. 
We believe (b) is more difficult to address and attempt some possible solutions. 

{\it {\bf Lazy Update -- Refreshing Enabling Block through Timeout:}} One way to address (b) is to have a small timeout value ($TO$) for $\cS_C$, 
which is inversely proportional to the turnover rate ($\zeta(n)$) of users in the system, i.e., $TO \propto 1/\zeta(n)$. 
The turnover rate is the ratio of the revoked users to all users, per unit of time.
This will enforce a small time window in which a revoked user can access the data, after which the routers caching the 
enabling block will expunge it. 
Any subsequent request for contents would require a fetch of the latest/updated enabling block. 

{\it {\bf Proactive Update -- Enabling Block refreshed by the CP:}} 
Another approach is the CP pushing the enabling block network-wide. 
Given that the number of users could be as several million spread across the globe, and that the 
data is cached at several hundred ISPs, this may not be very easy to accomplish.  
The challenges notwithstanding, such a proactive approach may be feasible with close interactions between 
the CP, the CDNs, and the ISPs. 

{\it {\bf Proactive Update -- Refreshing Enabling Block through Clustering:}} 
An improved approach is to partition the network into independent clusters with number of users 
$n^{\prime} < n$, where the clustering is motivated by access policies, geographical distribution, or cluster size. 
Each cluster $C_i$ has a cluster head (CH), which may be a CDN node or an ISP node, designated by the CP. 
Each $C_i$, uses a different polynomial $p_{t^{\prime}}^{i}(x)$, and given that $n^{\prime}$ could be smaller 
than $n$, the threshold $t^{\prime}$ can also be smaller than $t$. 
The enabling block may be generated at the CH or at the CP. 
In the event of a user revocation, now there is need for only a local update of the local $\cS^{i}_C$ corresponding to the cluster $C_i$. 
Updating the routers within the cluster during user revocation becomes much easier. 
%
The use of the smaller $t^{\prime}$ instead of $t$ will also speed-up the user's extraction procedure.  
A combination of the timeout and the clustering mechanisms may work better than either. 
%

\subsubsection{Number of Revoked Users Close to or Greater than $t$}
\label{sec04_02_02}   
There are two approaches to address this concern:
%

\noindent {\em Proactive Approach:} The CP can re-key the whole system with a new polynomial, 
and treat the already revoked users as non-existent; in essence re-initializing the system.
This procedure can be performed when the number of revoked users gets close to $t$. 

\noindent {\em Reactive Approach:} 
%
Let's consider the case where the number of revoked users $|\cR| = at+p << n$, where $a > 0$ and $p < t$.   
%
%
Let's assume that the key $\Upsilon$ is $128$-bits. 
As we mentioned in Section~5.B.2, the server splits $\Upsilon$ into $m$ pieces, $\{\tau_1, \ldots, \tau_m\}$. 
%
To ensure that revoked users cannot obtain $\Upsilon$, we can update Protocols~\ref{proto2} and~\ref{proto4} by choosing $t$ revoked users for 
each $\tau_i \in \Upsilon$ from $at+p$ revoked users, such that each revoked user $u_r$'s share is in the server share for at least one $\tau_i$.  
Then $u_r$ cannot decrypt one or more $\tau_i$s and hence $\Upsilon$ correctly. 
This can extend the scheme beyond $t$ revoked users. 

\vspace{-0.1in}
\subsection{System Reaches User Capacity}
\label{sec04_02_03} 
The system reaches user capacity when $n+t=\bZ^{*}_Q$.
At that point, no new users can be added.
All $x$s are allocated to users, no new unique user share can be created.
There is some scope for reuse of the tuples, with the initial revoked users' tuples replaced in the server's share. 
However, eventually the whole system has to be reinitialized with new prime numbers $Q' (>>Q)$ and $P' = 2Q'+1$, 
polynomial, and user tuples and distribution of the new user tuples and enabling blocks. 
However, we note that this would happen rarely.

\vspace{-0.05in}
\section{A Discussion on Security Provisions in AccConF}
\label{sec05b}
The security concerns in our framework include Sybil attacks, collusion attacks, and the 
other well-known attacks, such as CPA, CCA, and A-CCA. 
We will discuss how the framework can address these concerns. 
%
%
Unfortunately, in an ICN architecture, where routing is based on named data rather than 
hosts identifiers, there is no way to stop an impersonation attack or a Sybil attack. 
This is because, if a legitimate user is colluding with an impersonating user (sharing keys, passwords, etc.), then the impersonator has the 
keying materials of the legitimate node and can decrypt received content. 
The Sybil attack also follows similar reasoning. 
As pointed out by Douceur~\cite{Dou02}, it is difficult to handle such attacks without a central verification entity. 

{\em A possible way to identify an impersonating/Sybil node is by the server/CP requiring the user's player to periodically verify 
its credentials to the server.} 
The verification procedure can involve the CDN node and/or the ISP. 
Approximate location information obtained during these verifications (from CDN/ISP) can help estimate the user's geographic location 
(lower the entity in the hierarchy, the finer the localization). 
A user appearing at multiple locations simultaneously or over a short time span may be part of a 
Sybil or impersonation attack, and can be revoked. 
The clustering approach can further limit the impact of the attack. 
If each cluster uses a different polynomial, then a Sybil attacker using credentials of a user in a different cluster 
cannot decrypt the data.

A set of colluding nodes can create a new share for a new malicious (illegitimate) node, however this requires 
at least $t+1$ malicious/revoked nodes to collude, armed with the knowledge of $Q$ or $P$, so that they can re-generate 
the polynomial using their shares. 
With $t+1$ being of the order of thousands (or millions), this is unlikely. 
Note that the enabling block sent by the server cannot be used to obtain the legitimate shares as it is as hard as 
the DDH problem. 
For addressing the other attacks, such as CPA, CCA, and A-CCA, we refer the readers to~\cite{TzeTze01} -- the proofs are 
similar and we omit them here for brevity.      
%

Privacy of the users in an ICN is an important issue, with several privacy threats identified in the 
literature~\cite{ChaAbdCri12,lauinger12}. 
%
The most likely privacy threat is that of cache access monitoring, where an attacker connected to the same router as $u_i$ monitors the  
cache accesses of $u_i$ to track his behavior.
Even though in ICN, especially NDN, user's identity is not present in the packet, an attacker can leverage partial knowledge 
about the user (e.g., $u_i$ is interested in Sci-Fi movies) and the interest name to conjecture $u_i$'s identity. 
%
%
This problem is being studied by researchers~\cite{MohZhaSch13} and is not in scope for this paper. 
%
We note that if the secret key ($\tau$) is compromised, by means of some attack (a probable event in any secure system), 
it would require the content to be encrypted with a new key ($\tau'$).
Then the enabling block will also need to be updated according to the new key.

%% file: sec05.tex
\begin{figure}[t]
\centering
\hspace{-0.15in}
\subfigure[{Polynomial and Users Share Generation time (global)}]{
\label{fig:05_01a}
\includegraphics[width=1.7in]{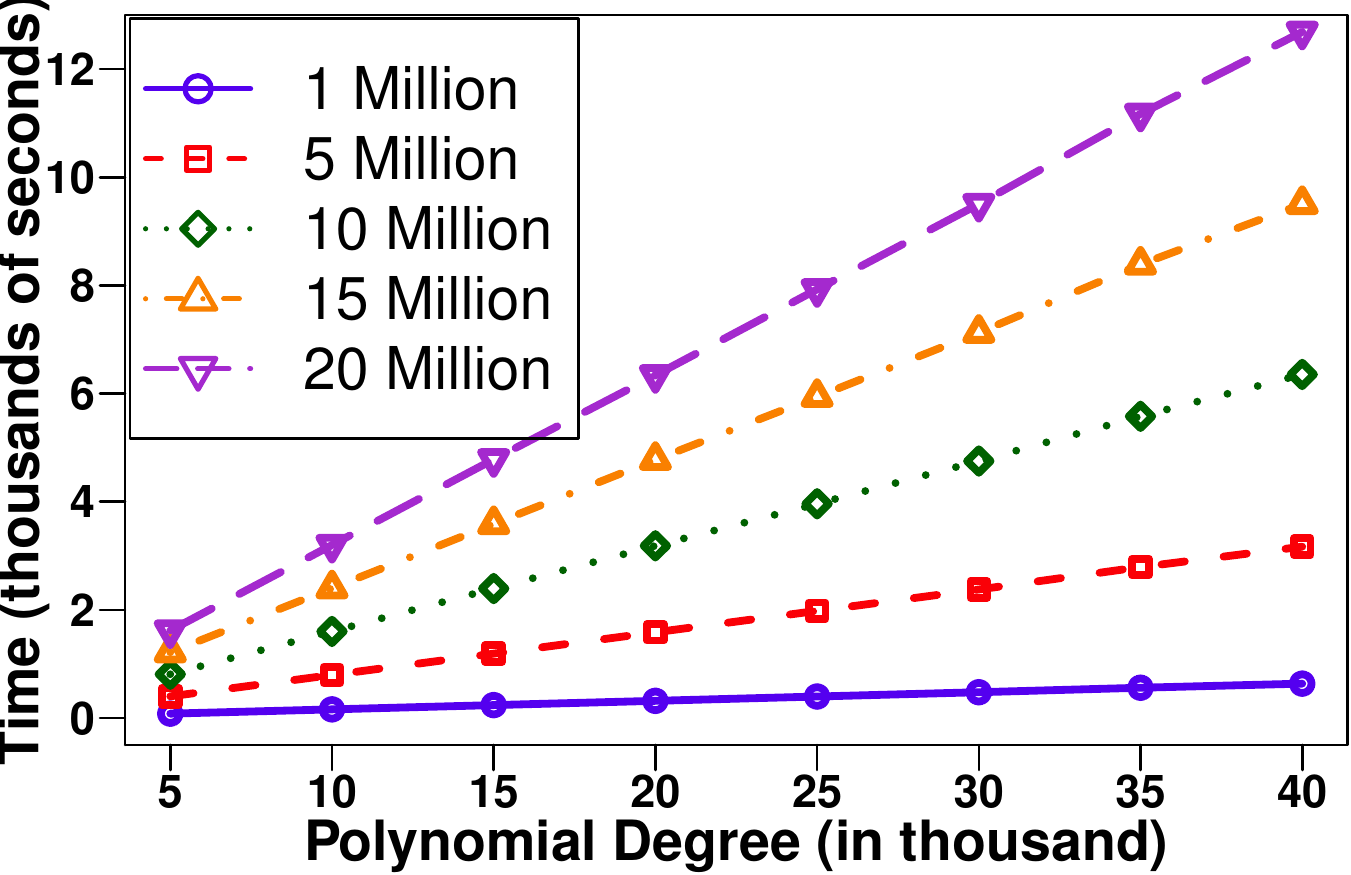}}
\hspace{-0.07in}
\subfigure[{Polynomial and Users Share Generation time}]{
\label{fig:05_01d}
\includegraphics[width=1.7in]{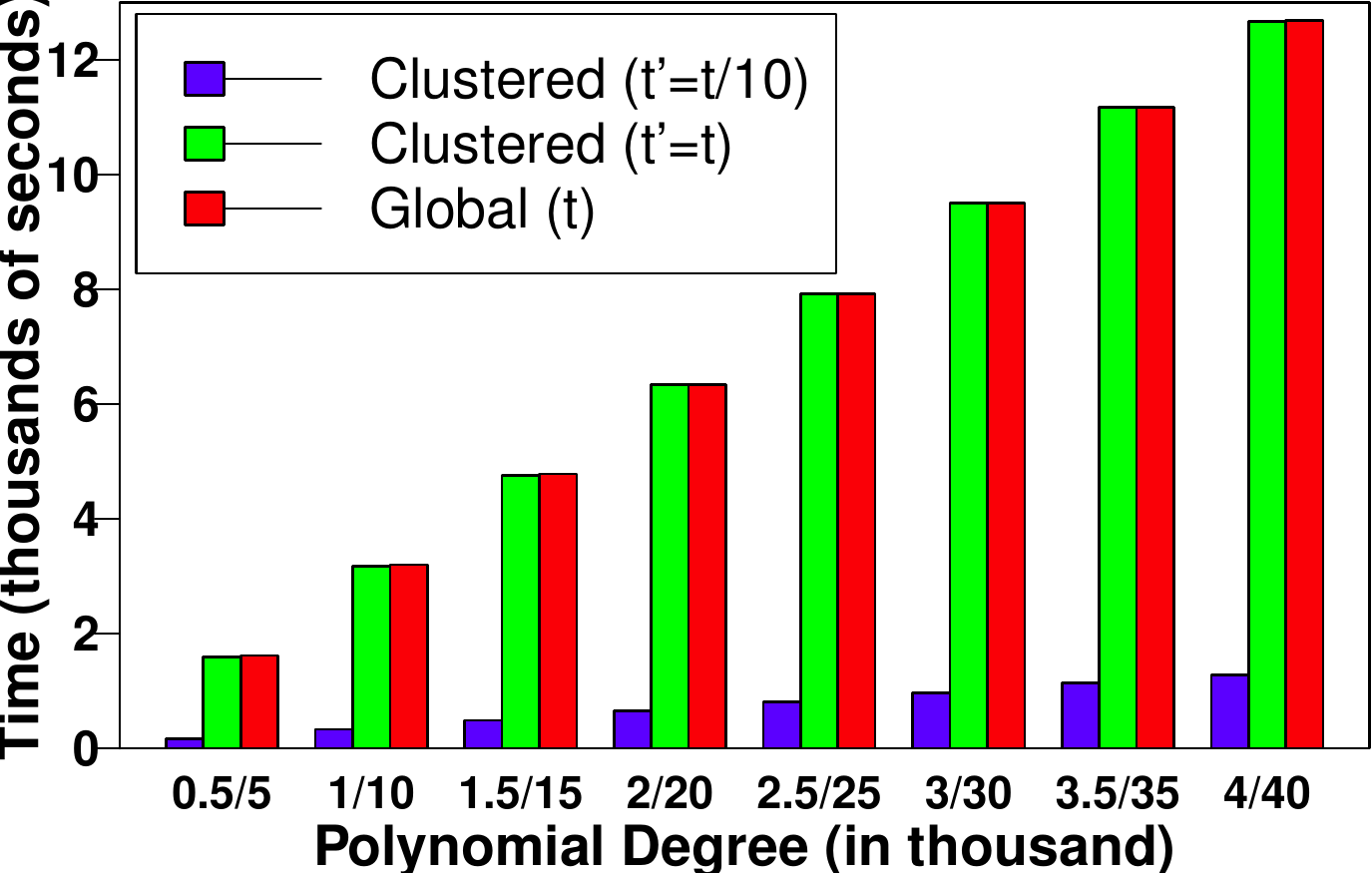}}
\hspace{0.1in}
\subfigure[Enabling Block Size]{
\label{fig:05_01e}
\includegraphics[width=1.65in]{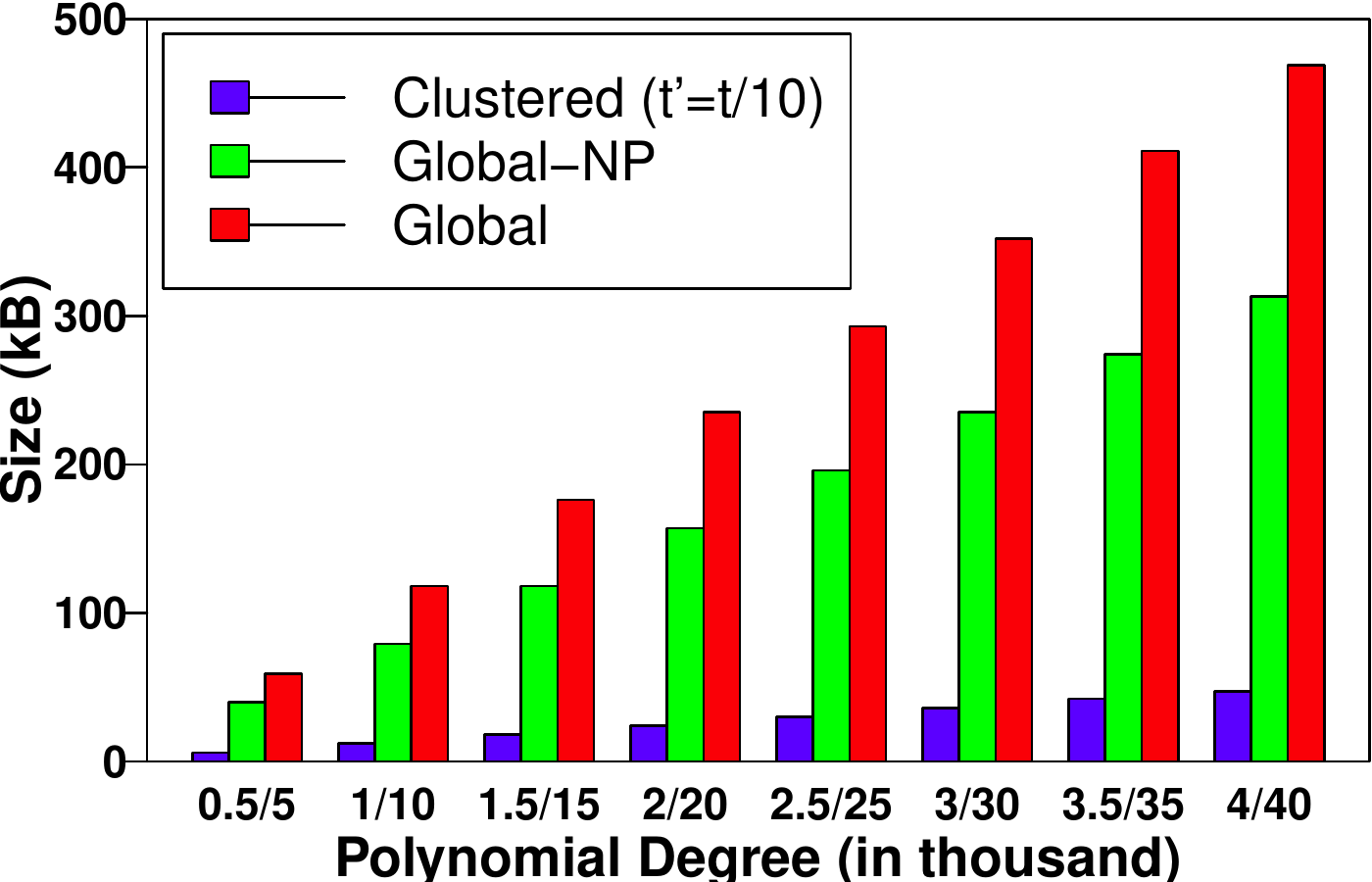}}
\hspace{-0.07in}
\subfigure[Symmetric Key Extraction]{
\label{fig:05_01f}
\includegraphics[width=1.65in]{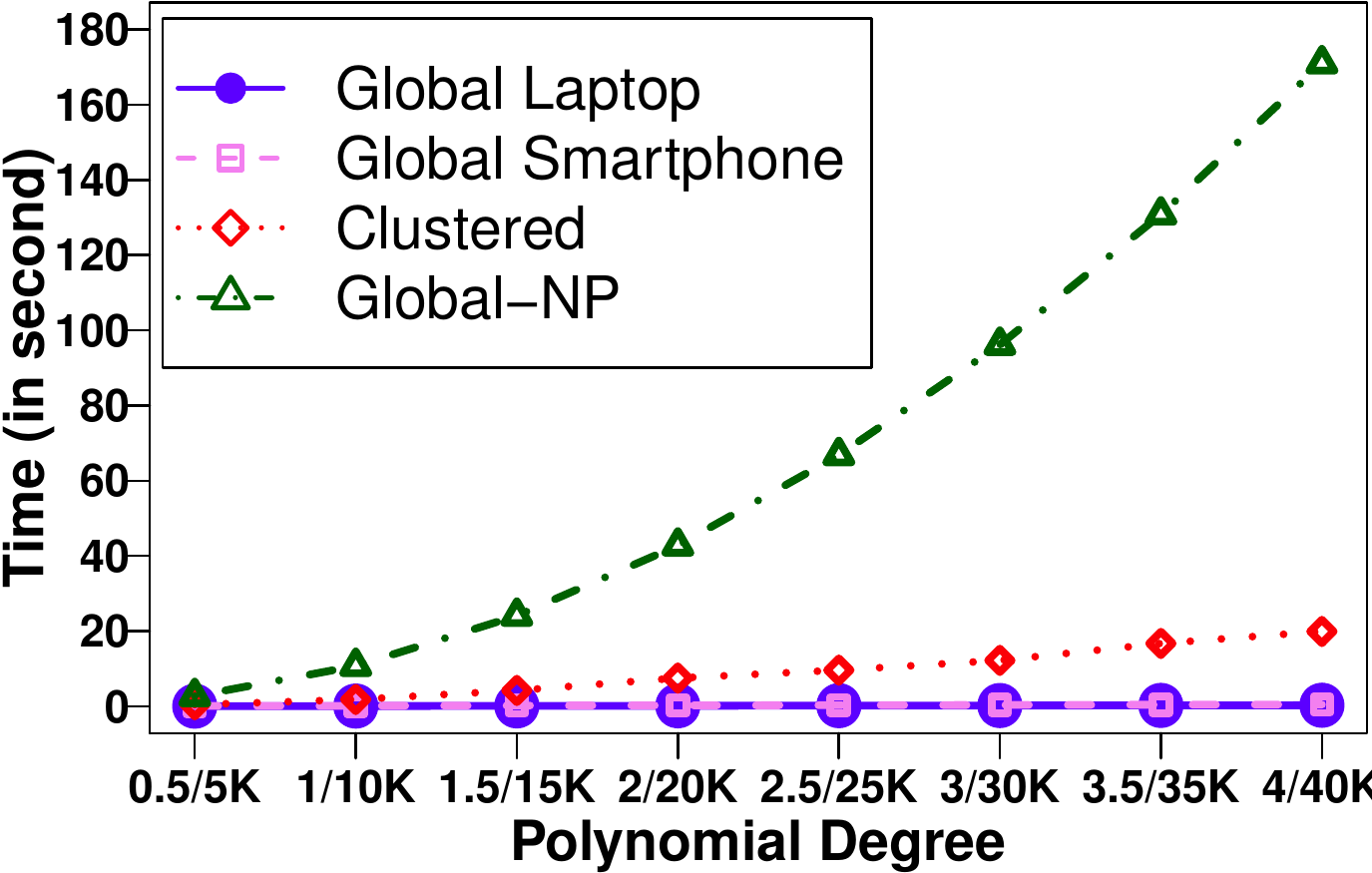}}
%
%
%
\caption{Results from Protocols Implementation: (a) Time taken to generate $p_t(x)$, $5K\leq t \leq 40K$ (global) and the user shares; 
(b) Comparison of time taken to generate $p_t(x)$, $5K \leq t \leq 40K$ and user shares in the global and 
two clustered scenarios ($5K\leq t^\prime (=t) \leq 40K$ and $0.5K\leq t^\prime (= t/10) \leq 4.0K$ respectively); 
(c) Size of the enabling block for the global and the clustered scenarios ($t^\prime = t; t^\prime = t/10$);  
(d) Time required for secret key extraction in the global and the clustered scenarios ($t^\prime = t; t^\prime = t/10$).} 
\label{fig:05_01}
\vspace{0.1in}
\end{figure}
%
\section{Implementation Results and Analyses}
\label{sec05}
%
Our implementation results are categorized into three segments: {\it (a)} experiments for performance analysis of our protocols with 
different settings; {\it (b)} experiments to assess the cost incurred for providing security (in terms of time) using AccConF over NDN; 
and {\it (c)} results from simulation using ndn-SIM on ns-3.
Our experiments were implemented on servers, laptops, and Nexus 5 smartphones. 
All these nodes were a part of a CCNx-0.7~\cite{Ccnx} testbed. 
%
%
%
%
%
For the first segment, on the laptop, we implemented our protocols in C (gcc version~4.5.2) and used the GNU 
Multi-Precision Arithmetic (GMP) library~\cite{Gmp} for cryptographic operations.
On the smartphones we used Android OS version 5.0.0 (Lollipop) and implemented the application using the Java based Android SDK API-19 (rev. 22.3) 
Kit Kat and NDK (rev. 9c).
Our mobile version was multithreaded and it decrypted the downloaded secret key $\tau$ concurrently while receiving content-chunks.

We implemented the Polynomial Generation protocol (Protocol~\ref{proto1}), the Enabling Block 
Generation and Encryption protocol (Protocol~\ref{proto2}), and the Extraction protocol (Protocol~\ref{proto4}). 
The straightforward user registration protocol was not implemented. 
%
%
In our implementation, for the global scenario, the total number of users ranged from $1$M to $20$M in 
increments of $5$M, and the value of $t$ ranged from $5$K to $40$K in increments of $5$K, where $M$ and $K$ 
stand for million and thousand respectively.
%
We chose $n \leq 20$M to represent the dynamic user base of a CP such as Netflix (by current estimates 
Netflix has $\approx 45$M users)~\cite{NflxUB}.
For the clustered scenario, there were 10 clusters, each having $2$M users; we assumed two sets of revocation 
thresholds ($t^\prime$): $t^\prime = t$ (as in global) and $t^\prime = t/10$, which ranged from $0.5$K to $4$K in increments of $0.5$K. 
%
Protocols~\ref{proto1} and~\ref{proto2}, were run on a server class machine with 24 Intel Xeon 2.40 GHz processors and 50~GB RAM. 
Only one processor was used in our experimental result.
Results were averaged over 100 runs.
%

%
%
Fig.~\ref{fig:05_01} displays the results for polynomial and enabling block generations and key extraction. 
Fig.~\ref{fig:05_01a} shows the time taken to generate polynomials of different degrees, consisting of generating 
random coefficients ($\{a_0, \ldots, a_t\}$) for the polynomial ($p_t(x)$), and then evaluating $p_t(x)$ at $n + t$ points. 
%
%
The $X$-axis represents different polynomial degrees (equivalent to $t$) and the $Y$-axis represents the time in thousands of seconds. 
%
%
%
{\em The polynomial generation procedure is the most time consuming component of our framework, however, it is 
executed by the server only and can be performed offline and in parallel by several processors.} 
We note that the increase in running time with increasing $t$ (for different values of $n$) is attributable not 
only to the polynomial degree but also the number of users. 
The running time scales linearly---the generation time for $20$M users is $20$ times 
more than that for $1$M users ($t$ being the same). 
%

Fig.~\ref{fig:05_01d} shows a comparison between the two clustered scenarios ($t^{\prime}=t$ and $t^{\prime} = t/10$) and the global 
scenario on the basis of the polynomial generation time. 
When $t^\prime = t$, as expected, the time taken is the same. 
%
%
%
%
For the $t^{\prime}=t/10$ case, running time for one polynomial generation is obviously going to be small. 
Interestingly, the running time for generating the ten polynomials of degree $t^{\prime}$ is much less than generating a 
polynomial of degree $t$.
%

%
Fig.~\ref{fig:05_01e} shows the size of the enabling block $\cS_C$ in the clustered scenario $t^\prime = t/10$ and 
two different global scenarios: one in which AccConF is used, thus the partial Lagrangian coefficients are precomputed at 
the server (Protocol~\ref{proto2}), denoted as {\em Global}, and the other in which no precomputation is performed at the 
server, denoted as {\em GlobalNP}. 
%
The X-axis represents the polynomial degree ($t$) and the Y-axis represents size in KiloBytes. 
%
The size of $\cS_C$ is independent of the number of members; it increases proportionally with $t$. 
%
The worst case size is in Global $\approx470$kB, for $t= 40$K. 
Even then, given that a standard two-hour Netflix movie on a mobile device has a size of around $300$ MB~\cite{MBw}, the 
enabling block makes-up less than $0.16$\% of the movie!  
%
%
%
For the reactive approach, to handle the case where the number of revoked user $|\cR|> t$, the size of the 
enabling block will increase.
%
Even if we create a share for every byte of a $128$-bit secret key (allows $16 \cdot t$ revocations), 
the enabling block size is $7.52$MB--only a $2.5\%$ overhead.
%

The extra precomputed information at the server results in the enabling block in Global to be significantly  
more than that of GlobalNP (around $50$\% for $t = 40$K). 
However, as shown in Fig.~\ref{fig:05_01f}, the {\em corresponding reduction in extraction time at the client due to 
the precomputation (Global) is significant}--less than 1~second for both the laptop and smartphone versions.
The key extraction time in GlobalNP grows super-linearly with increasing $t$. 
The laptop client was running on an Apple Macbook Pro running VMware, allocated 1~GB RAM and one 2.5~GHz, Intel Core~i5 processor. 
The smartphone was a Nexus 5, 2.3~GHz quad core, 2~GB smartphone.  
%
%
%
%

As a demonstration of the framework's scalability, especially from the perspective of the expensive extraction protocol, 
we obtained statistics for higher values of $t$ as shown in Table~\ref{table2}.   
Even when $t$ is $1$ million, the enabling block is only $12$ MB ($1.2$\% of a standard Netflix movie~\cite{NflxUB}) 
and the corresponding extraction of $\tau$ takes $1.34$~seconds on the laptop and $10.65$~seconds 
on the smartphone.  
The difference between the laptop and smartphone results are due to the difference in their processors (smartphone's low-power 
processors are slower). 
Also, in the laptop the algorithms are implemented in C, while on the smartphones they run on the Java based Android SDK. 
%
%
%
%
%

\setlength{\textfloatsep}{0.05cm}
\begin{table}[!t]
\centering
\caption{Statistics for Large Values of $t$ related to Extraction of $\tau$}
\label{table2}
\begin{tabular}{|p{1.42in}|c|c|c|c|c|}
 \hline
 	\raggedright $t$ (in million) & $0.1$ & $0.3$ & $0.5$ & $0.7$ & $1$ \\
 \hline
	\raggedright Laptop Extraction Time (secs) & $0.14$ & $0.46$ & $0.71$ & $1.03$ & $1.34$ \\ 
 \hline
	\raggedright Smartphone Ext. Time (secs) & $1.16$ & $3.68$ & $5.92$ & $7.44$ & $10.65$ \\ 
 \hline
	\raggedright Enabling Block Size (MB) & $1.2$ & $3.6$ & $6$ & $8.4$ & $12$ \\
 \hline
  	\raggedright Smartphone RAM Usage (MB) & $12$ & $40$ & $70$ & $96$ & $143$ \\
 \hline
\end{tabular}
\end{table}

Revocation threshold of $1$ million is large, as can be seen from recent Netflix statistics~\cite{NflxUB,PARK},  
reached on an average in three months for Netflix.  
This makes system re-initialization events rare and scalable.
%
%
Eventually, a successful/scalable implementation should combine our clustering approach and 
smaller values of $t$ (say $100,000$), which will allow a smartphone to extract $\tau$ in close to $1$~second. 
Such implementations can handle values of $n$ close to $1$ million in a cluster. 
Also, our approach has a modest memory footprint, the high RAM usage numbers (e.g., $143$~MB for $t=1$~million) are only during the 
extraction process. 
%

\begin{figure}[b]
\centering
\includegraphics[width=2.8in]{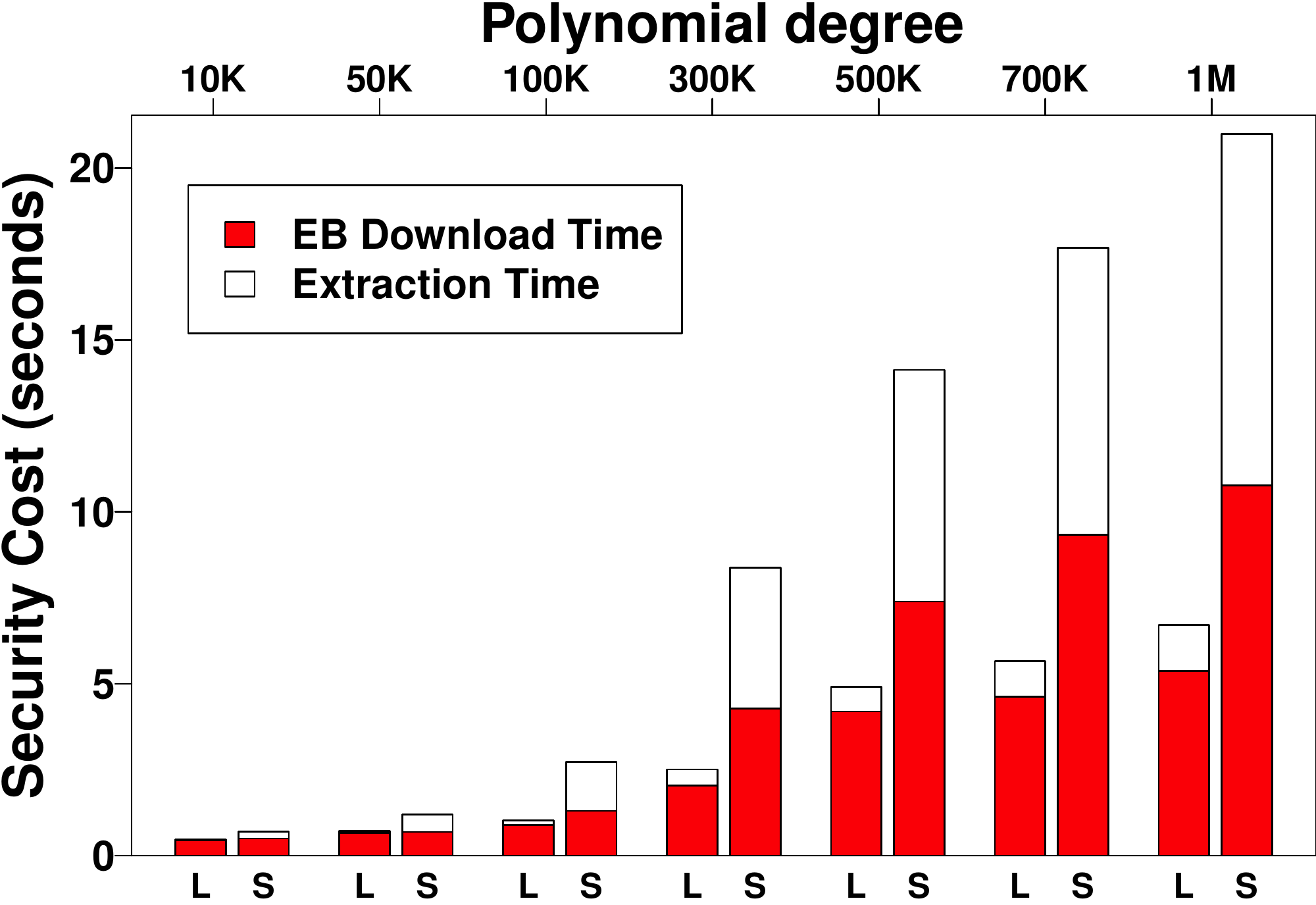}
\caption{Security Cost for the Laptop (L) User and the Smartphone (S) User.}
\label{fig:05_02}
\end{figure}

%
For the second results segment, we implemented one client on the Macbook, the smartphone version on the Nexus 5 and the CP 
(a server with $2.5$~GHz Intel Core 2 Quad, $3.8$~GB) was five hops away from the clients 
over a four-tiered network (created using switches and IPv4 routers). 
We compared the baseline NDN's and AccConF's performance in content retrieval.
Our framework took almost the same time for content download, the additional delay being in downloading the enabling block 
and extracting the secret key---an overhead to enforce the AC.
Hence, we define the security cost as the total extra time that it takes for a client to download the enabling block from a 
nearby cache and extract the secret.
Fig.~\ref{fig:05_02} illustrates AccConF's security cost, for different polynomial degrees, for the laptop and the smartphone clients.  
\vspace{0.08in}
The cost for the smartphone application increases faster than the laptop's; this can be attributed to the better resources at the laptop's 
disposal. 
The biggest cost for the laptop is downloading the enabling block, whereas in the smartphone the costs of communication and 
extraction are almost comparable. 
It is interesting that the download time for the smartphone is higher than the laptop despite both connecting to the 
same access point using IEEE 802.11n. 
This difference is attributable to the laptop antenna being more powerful than the smartphone antenna. 

Our last segment details our simulation results using ndn-SIM on ns-3. 
We simulated the AccConF, NDN, and the UDP clients on ten different network topologies; 
we illustrate the results of four representative ones.
The four representative network topologies were: \{$3755$ nodes, $7449$ edges\}; \{$3709$ nodes, $7193$ edges\}; 
\{$3707$ nodes, $7353$ edges\}; and \{$3696$ nodes, $7331$ edges\}. 
%
%
The topologies were created using the two-layer Top-Down hierarchical model in BRITE~\cite{brite}. 
The autonomous systems (AS) layer was created using the Waxman model and the router layer for each AS was created 
using the Barab\`{a}si-Albert model.
Each topology had two edge routers, each serving five clients through $20$Mbps links. 
One content provider was placed across the network, $6$ to $8$ hops from the two edge routers. 
Links in the network core had bandwidth selected randomly between $1$ to $4$~Gbps. 

The server contained $100$ content objects. 
Each object was $300$~MB for NDN and UDP and $312$~MB for AccConF ($12$~MB for the enabling block) respectively. 
The content popularity followed a Zipf-Mandelbrot distribution with $q=1$ and $s=2$, which is reflected in the requests 
made by the clients. 
The clients constantly requested content--if one content request was satisfied they requested another. 
For fair comparison, the chunk size was $1436$~bytes in NDN and AccConF, comparable to a standard Ethernet frame size. 
In NDN and AccConF, the routers were equipped with $1.5$~GB cache (i.e., $5\%$ of the entire content) and used 
the LRU cache-eviction algorithm.
{\it {We ran the simulation for 30000 seconds.}}
The simulations were run on a server-class machine having 2~AMD Opteron G276 processors, each core clocking $2.3$~GHz, 
with $128$~GB~RAM.

\begin{figure}[h]
\centering
\subfigure[Topology~1]{
\label{fig:07_04a}
\includegraphics[width=1.0in]{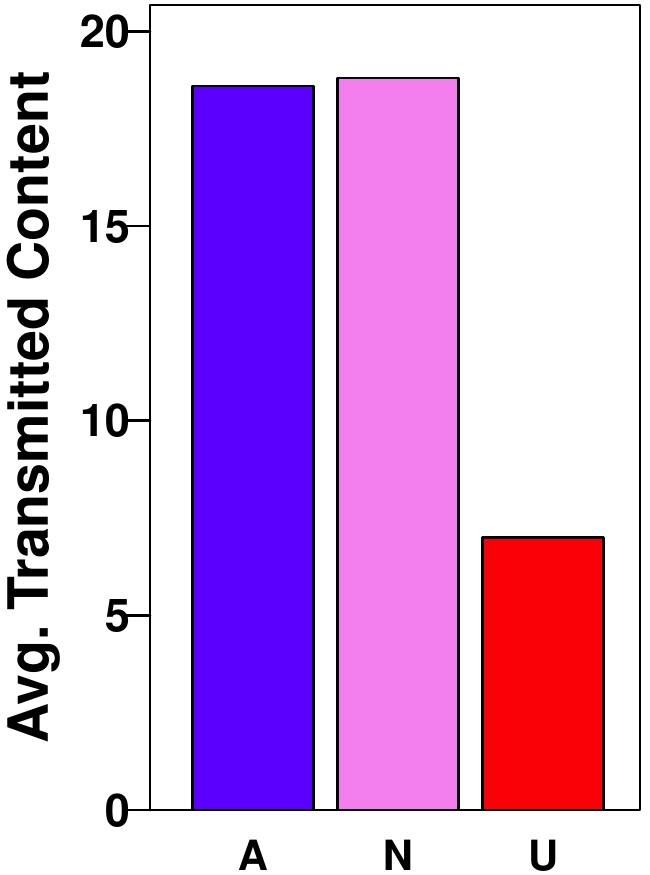}}
\hspace{0.25in}
\subfigure[Topology~2]{
\label{fig:07_04b}
\includegraphics[width=1.0in]{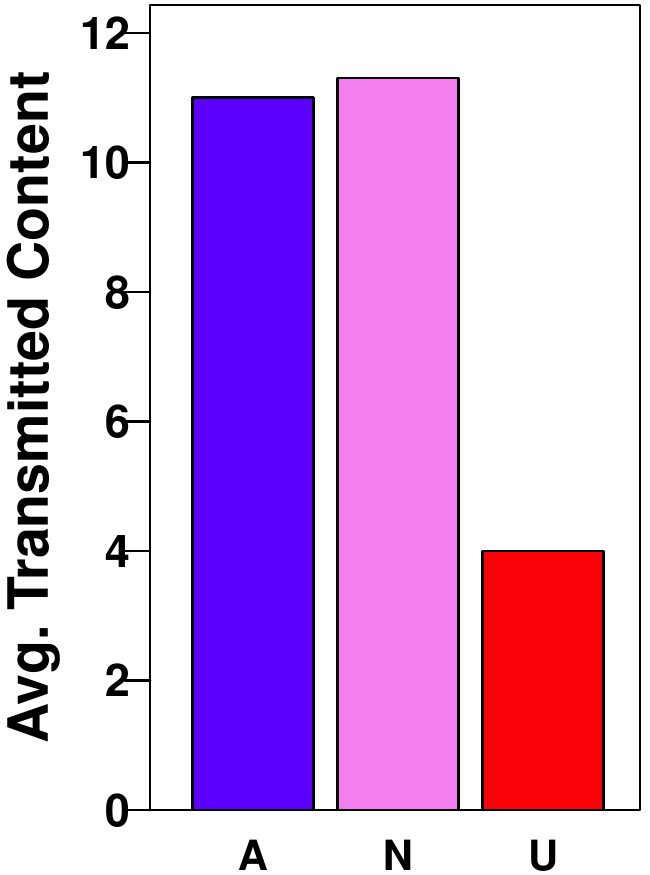}}
\subfigure[Topology~3]{
\label{fig:07_04c}
\includegraphics[width=1.0in]{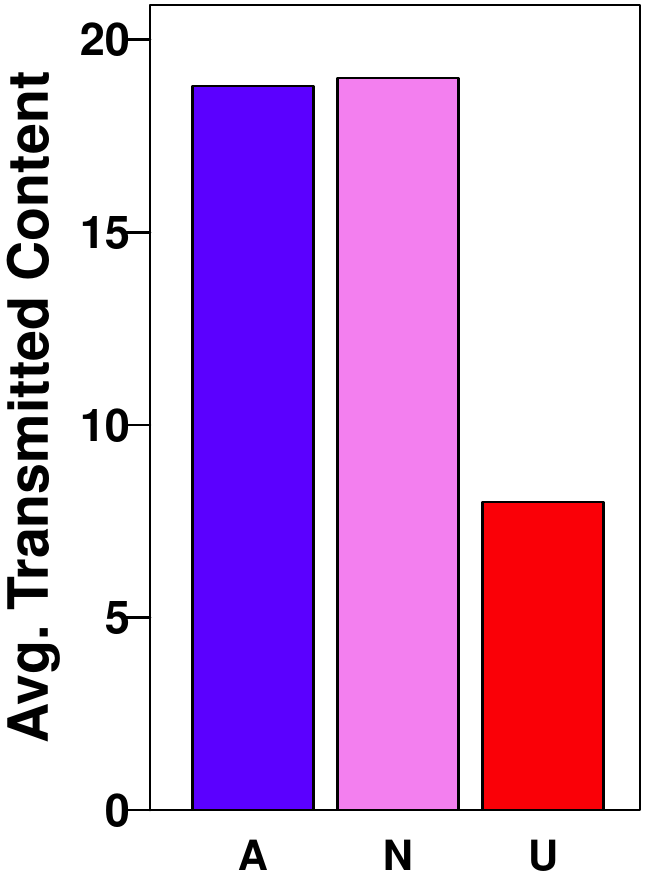}}
\hspace{0.25in}
\subfigure[Topology~4]{
\label{fig:07_04d}
\includegraphics[width=1.0in]{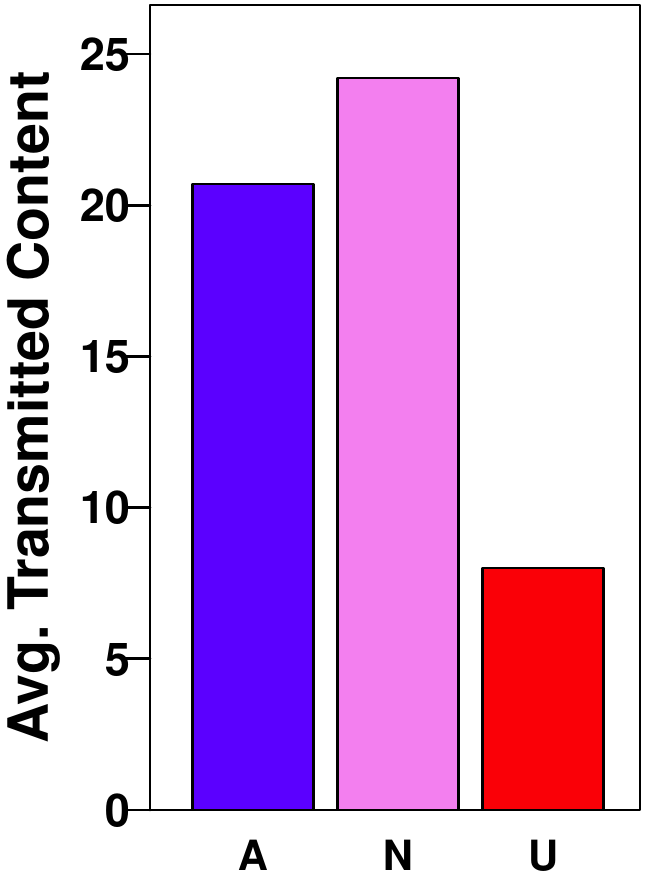}}
\centering
\vspace{-0.1in}
\caption{Average Number of Content Downloaded per client in AccConF (A), NDN (N), and UDP (U).}
\label{fig:07_04}
\end{figure}

Fig.~\ref{fig:07_04} shows the average number of contents downloaded by each client. 
In NDN and AccConF, the clients' requests are satisfied faster by virtue of nearby caches, hence the clients request 
more contents. 
NDN performs a little better than AccConF because it does not have the enabling block. 
In the last topology, the margin is relatively wider. 
%
From our analysis, we identified that this topology's  structure is such that more requests are completed, which leads to 
more cache-evictions and hence more requests being served from farther caches. 
Consequently AccConF is punished more on account of its enabling block overhead. 

\begin{figure}[!tph]
\centering
\subfigure[Topology~1]{
\label{fig:05_04a}
\includegraphics[width=3.3in,height=1.1in]{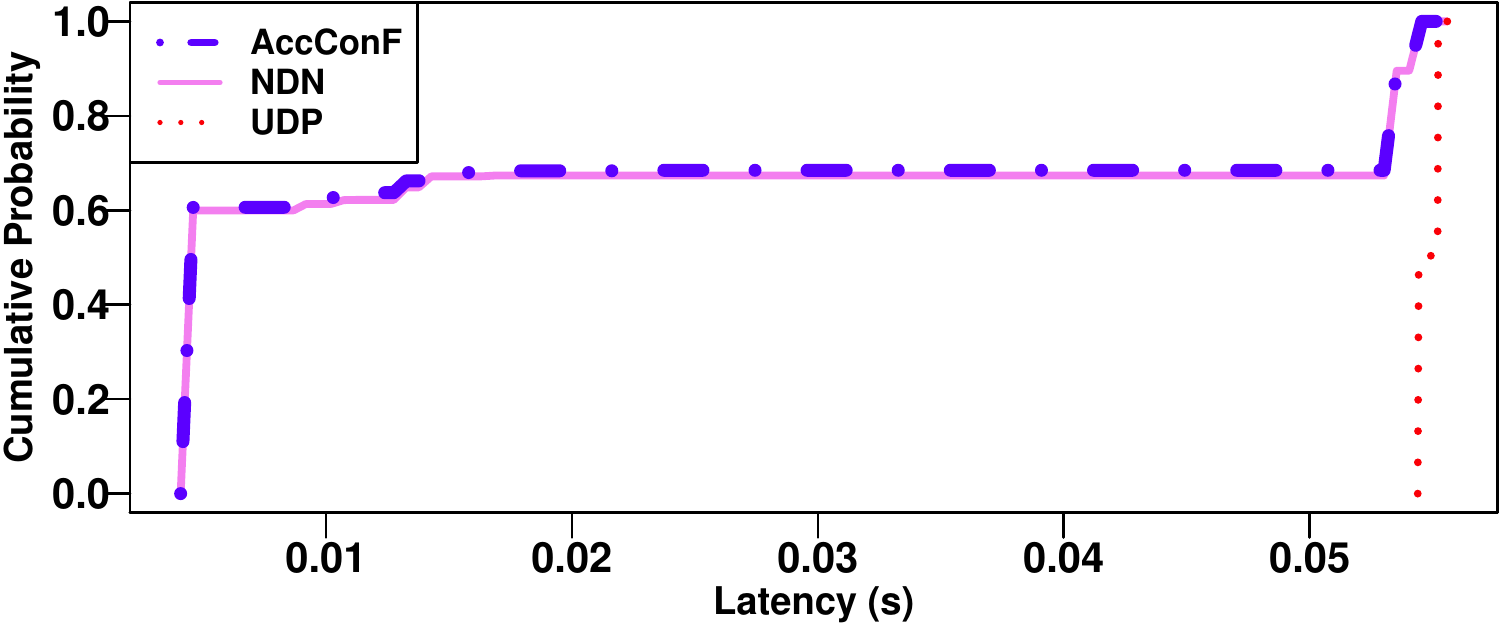}}
%
\subfigure[Topology~2]{
\label{fig:05_04b}
\includegraphics[width=3.3in,height=1.1in]{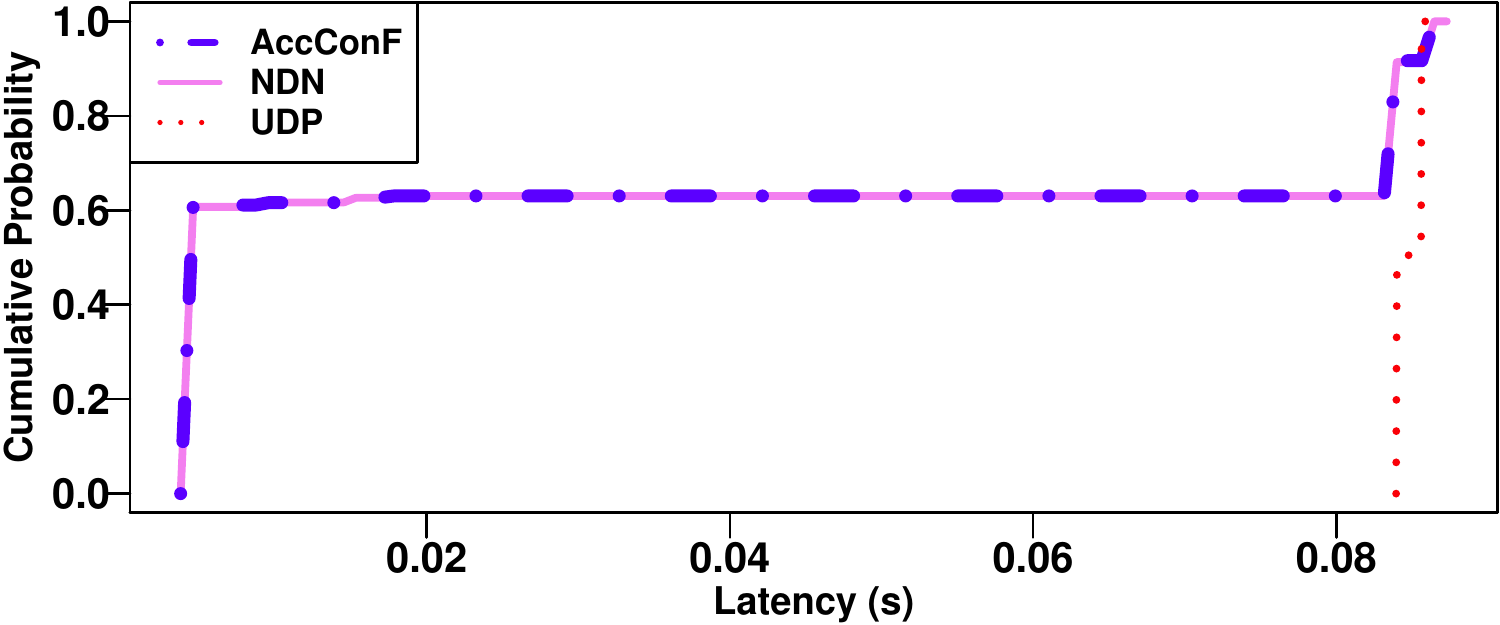}}
\subfigure[Topology~3]{
\label{fig:05_04c}
\includegraphics[width=3.3in,height=1.1in]{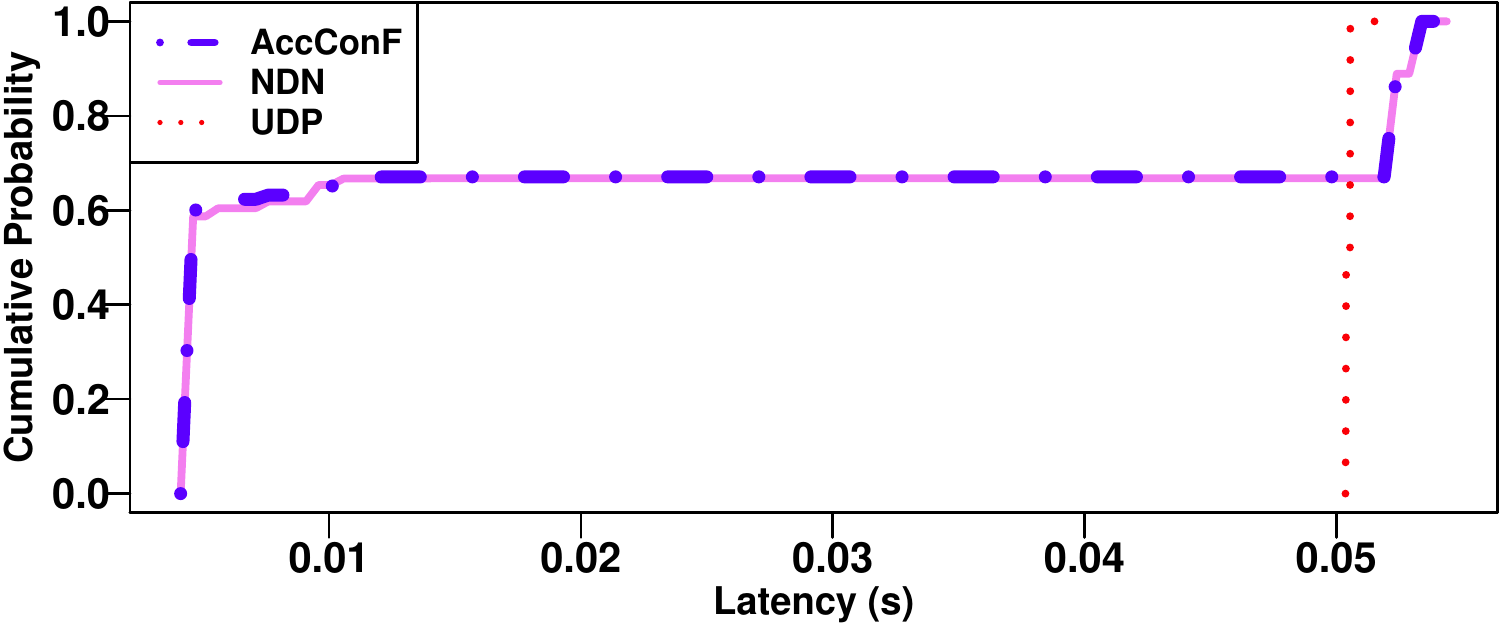}}
%
\subfigure[Topology~4]{
\label{fig:05_04d}
\includegraphics[width=3.3in,height=1.1in]{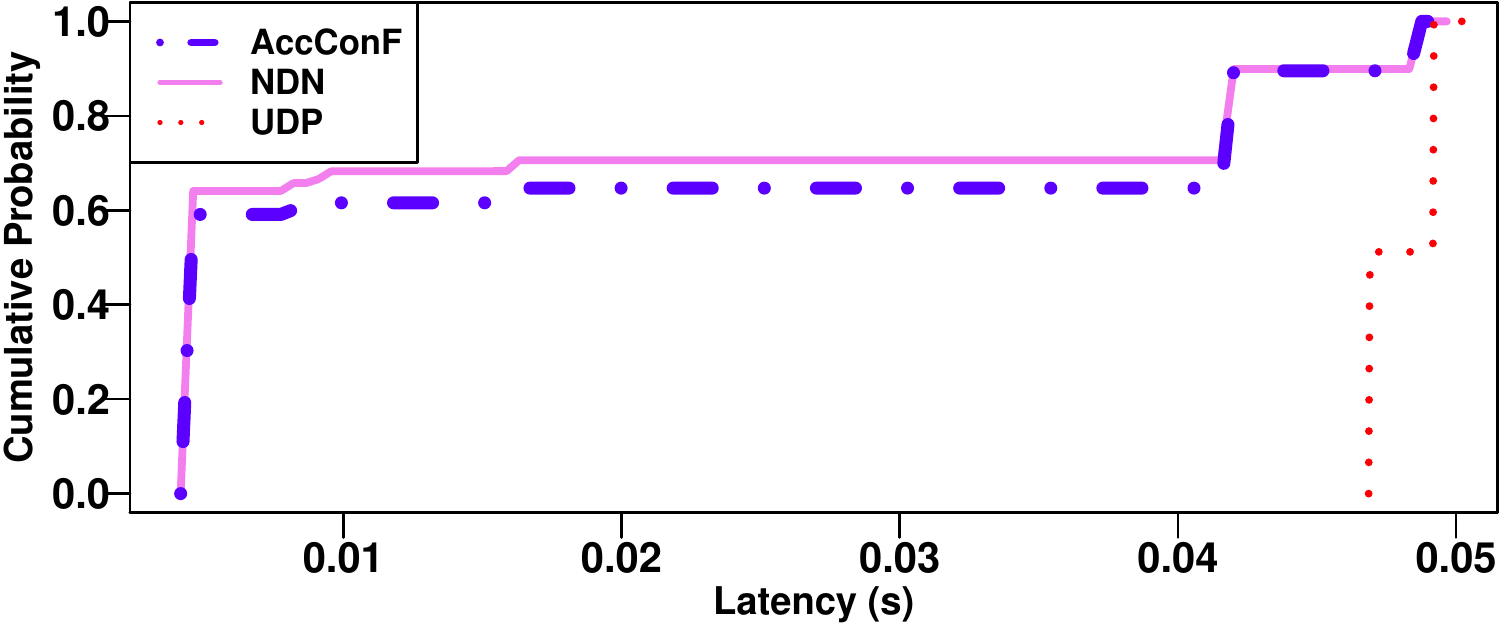}}
%
\caption{eCDF for the Latency in AccConF (A), NDN (N), and UDP (U).}
\label{fig:05_04}
\end{figure}

Fig.~\ref{fig:05_04} presents the empirical cumulative distribution function (eCDF) for per-interest latencies in the three approaches. 
NDN and AccConF have a significant number of interests that are served in less than $0.01$ seconds, which markedly improves the 
number of contents downloaded. 
Fig.~\ref{fig:05_04d} further illustrates why AccConF has lesser per-client downloads. 
Whereas in the first three topologies the cumulative probabilities of AccConF and NDN track closely, here AccConF 
is served by a farther cache (reflected in the eCDF increasing after latency value of $0.042$s).
%
%
%

%% file: paper.bbl
\begin{thebibliography}{10}

\bibitem{AhlDanImb12}
B.~Ahlgren, C.~Dannewitz, C.~Imbrenda, D.~Kutscher, and B.~Ohlman.
\newblock A survey of information-centric networking.
\newblock {\em IEEE Communications Magazine}, 50(7):26--36, 2012.

\bibitem{AriKop12}
S.~Arianfar, T.~Koponen, B.~Raghavan, and S.~Shenker.
\newblock On preserving privacy in content-oriented networks.
\newblock In {\em ACM SIGCOMM Information-centric networking (ICN) workshop},
  pages 19--24. ACM, 2011.

\bibitem{brite}
Brite: Boston university representative internet topology generator, 2014.
\newblock http://www.cs.bu.edu/brite.

\bibitem{ChaAbdCri12}
A~Chaabane, E~De~Cristofaro, M.~Kaafar, and E.~Uzun.
\newblock Privacy in content-oriented networking: Threats and countermeasures.
\newblock {\em arXiv preprint arXiv:1211.5183}, 2012.

\bibitem{CheLeiXu14}
T.~Chen, K.~Lei, and K.~Xu.
\newblock An encryption and probability based access control model for named
  data networking.
\newblock In {\em IEEE IPCCC}, pages 1--8. IEEE, 2014.

\bibitem{Cis2}
Cisco.
\newblock Cisco visual networking index forecast (2019), 2016.
\newblock {\sf
  http://www.cisco.com/c/en/us/solutions/service-provider/visual-networking-index-vni/vni-forecast.html}.

\bibitem{Dan09}
C.~Dannewitz.
\newblock {NetInf: An information-centric design for the future Internet}.
\newblock In {\em 3rd GI/ITG KuVS Workshop on The Future Internet}, 2009.

\bibitem{Dou02}
J.~Douceur.
\newblock The sybil attack.
\newblock {\em Peer-to-peer Systems}, pages 251--260, 2002.

\bibitem{FiaNao94}
A.~Fiat and M.~Naor.
\newblock Broadcast encryption.
\newblock In {\em CRYPTO}, pages 480--491, 1994.

\bibitem{FotMarPol12}
N.~Fotiou, G.F. Marias, and G.C. Polyzos.
\newblock Access control enforcement delegation for information-centric
  networking architectures.
\newblock In {\em ACM Information-centric Networking Workshop}, pages 85--90,
  2012.

\bibitem{FotNikTro10}
N.~Fotiou, P.~Nikander, D.~Trossen, and G.C. Polyzos.
\newblock {Developing information networking further: From PSIRP to PURSUIT}.
\newblock In {\em ICST Conference on Broadband Communications, Networks, and
  Systems}, pages 1--13, 2010.

\bibitem{GhaSchTsu15}
C.~Ghali, M.~Schlosberg, G.~Tsudik, and C.~Wood.
\newblock Interest-based access control for content centric networks (extended
  version).
\newblock {\em arXiv preprint arXiv:1505.06258}, 2015.

\bibitem{Gmp}
{The GNU Multiple Precision Arithmetic Library}, 2012.
\newblock {\sf http://www.gmplib.org}.

\bibitem{IonZhaSch13}
M.~Ion, J.~Zhang, and E.~M. Schooler.
\newblock Toward content-centric privacy in icn: attribute-based encryption and
  routing.
\newblock {\em ACM SIGCOMM Computer Comm. Review}, 43(4):513--514, 2013.

\bibitem{JacSmeTho09}
V.~Jacobson, D.K. Smetters, J.D. Thornton, M.F. Plass, N.H. Briggs, and R.L.
  Braynard.
\newblock Networking named content.
\newblock In {\em Intl. conference on Emerging networking experiments and
  technologies}, pages 1--12. ACM, 2009.

\bibitem{KopChaChu07}
T.~Koponen, M.~Chawla, B.~Chun, A.~Ermolinskiy, K.~Kim, S.~Shenker, and
  I.~Stoica.
\newblock A data-oriented (and beyond) network architecture.
\newblock {\em ACM SIGCOMM Computer Communication Review}, 37(4):181--192,
  2007.

\bibitem{Ccnx}
Palo Alto~Research Lab.
\newblock Ccnx.
\newblock {\sf http://www.ccnx.org/}.

\bibitem{lauinger12}
T.~Lauinger, N.~Laoutaris, P.~Rodriguez, and {\em et al.}
\newblock Privacy implications of ubiquitous caching in named data networking
  architectures.
\newblock Technical report, Technical Report TR-iSecLab-0812-001, iSecLab,
  2012.

\bibitem{LiVerHua14}
B.~Li, A.P. Verleker, D.~Huang, Z.~Wang, and Y.~Zhu.
\newblock Attribute-based access control for icn naming scheme.
\newblock In {\em IEE Conference on Communications and Network Security}. IEEE,
  2014.

\bibitem{LiZhaZhe15}
Q.~Li, X.~Zhang, Q.~Zheng, R.~Sandhu, and X.~Fu.
\newblock Live: Lightweight integrity verification and content access control
  for named data networking.
\newblock {\em IEEE Transactions on Information Forensics and Security},
  10(2):308--320, 2015.

\bibitem{MBw}
{App Makers Worry as Data Plans Are Capped}, June 6, 2010.
\newblock {\sf http://www.nytimes.com/2010/06/07/technology/07data.html?\_r=0}.

\bibitem{MenOorVan97}
A.J. Menezes, P.C. Van~Oorschot, and S.A. Vanstone.
\newblock {\em Handbook of applied cryptography}.
\newblock CRC, 1997.

\bibitem{MisTouMaj13}
S.~Misra, R.~Tourani, and N.~Majd.
\newblock Secure content delivery in information-centric networks: design,
  implementation, and analyses.
\newblock In {\em Proceedings of the ACM SIGCOMM workshop on
  Information-centric networking}, pages 73--78. ACM, 2013.

\bibitem{MohZhaSch13}
A.~Mohaisen, X.~Zhang, M.~Schuchard, H.~Xie, and Y.~Kim.
\newblock Protecting access privacy of cached contents in information centric
  networks.
\newblock In {\em ACM SIGSAC Symposium}, pages 173--178. ACM, 2013.

\bibitem{NaoNaoLot01}
D.~Naor, M.~Naor, and J.~Lotspiech.
\newblock Revocation and tracing schemes for stateless receivers.
\newblock In {\em CRYPTO}, pages 41--62, 2001.

\bibitem{NaoPin01}
M.~Naor and B.~Pinkas.
\newblock Efficient trace and revoke schemes.
\newblock In {\em Financial cryptography}, pages 1--20, 2001.

\bibitem{NflxUB}
{Netflix has over 69 million members in over 60 countries.}, October 25, 2011.
\newblock {\sf http://ir.netflix.com/}.

\bibitem{PARK}
{OTT Subscriber Annual Churn Rates.}, July 30, 2015.
\newblock {\sf
  https://www.parksassociates.com/blog/article/pr-july2015-ott-tracker}.

\bibitem{Sch91}
C.-P. Schnorr.
\newblock Efficient signature generation by smart cards.
\newblock {\em Journal of Cryptology}, 4(3):161--174, 1991.

\bibitem{TarAinVis09}
S.~Tarkoma, M.~Ain, and K.~Visala.
\newblock The publish/subscribe internet routing paradigm (psirp): Designing
  the future internet architecture.
\newblock {\em Towards the Future Internet}, page 102, 2009.

\bibitem{Tor}
{Tor Project: Anonymity Online}.
\newblock {\sf http://www.torproject.org/}.

\bibitem{TzeTze01}
W.~Tzeng and Z.~Tzeng.
\newblock A public-key traitor tracing scheme with revocation using dynamic
  shares.
\newblock In {\em Public Key Cryptography}, pages 207--224, 2001.

\bibitem{Wan11}
S.~Wang, J.~Bi, J.~Wu, Z.~Li, W.~Zhang, and X.~Yang.
\newblock Could in-network caching benefit information-centric networking?
\newblock In {\em 7th Asian Internet Engineering Conference}, pages 112--115,
  2011.

\bibitem{XieWidWan12}
M.~Xie, I.~Widjaja, and H.~Wang.
\newblock Enhancing cache robustness for content-centric networking.
\newblock In {\em IEEE INFOCOM}, pages 2426--2434, 2012.

\end{thebibliography}
